%% file: HQmDisJ.tex
\documentclass[12pt]{JHEP3}
\usepackage{graphicx,amssymb}
\usepackage{citesort}

\input{text/HQmDis.fig}

\title{ Heavy Quark Mass Effects in Deep Inelastic Scattering and
Global QCD Analysis }

\author
{W.K.~Tung\thanks{E-mail: tung@pa.msu.edu}\ $^{a,b}$, H.L.~Lai$^c$,
A.~Belyaev$^a$, J.~Pumplin$^a$,
D.~Stump$^a$, and C.-P.~Yuan$^a$ \\
$^a$ Michigan State University, E. Lansing, MI, USA \\
$^b$ University of Washington, Seattle, WA, USA \\
$^c$ Taipei Municipal University of Education, Taipei, Taiwan }

\abstract {A new implementation of the general PQCD formalism of
Collins, including heavy quark mass effects, is described. Important
features that contribute to the accuracy and efficiency of the
calculation of both neutral current (NC) and charged current (CC)
processess are explicitly discussed. This new implementation is
applied to the global analysis of the full HERA I data sets on NC
and CC cross sections, with correlated systematic errors, in
conjunction with the usual fixed-target and hadron collider data
sets. By using a variety of parametrizations to explore the parton
parameter space, robust new parton distribution function (PDF) sets
(CTEQ6.5) are obtained. The new quark distributions are consistently
higher in the region $x\sim10^{-3}$ than previous ones, with
important implications on hadron collider phenomenology, especially
at the LHC. The uncertainties of the parton distributions are
reassessed and are compared to the previous ones. A new set of
CTEQ6.5 eigenvector PDFs that encapsulates these uncertainties is
also presented.}

\preprint{ MSU-HEP-0605 }

\newcommand{\msbar}{\mbox{\small {$\overline {MS}$}}}

\begin{document}

\setcounter{footnote}{0}

\input{text/1.intro}
\input{text/2.implement}
\input{text/2.difference}
\input{text/3.globalfit}
\input{text/4.comparison}
\input{text/5.uncertainty}
\input{text/6.summary}
\appendix
\input{text/a1.rescale.tex}
\input{text/a2.param.tex}

\input{text/HQmDis.cit.tex}

\end{document}

%% file: text/HQmDis.fig.tex
\newcommand{\figRescaling}
{
\FIGURE[h]{
\resizebox*{!}{0.4\textheight}{
 \includegraphics{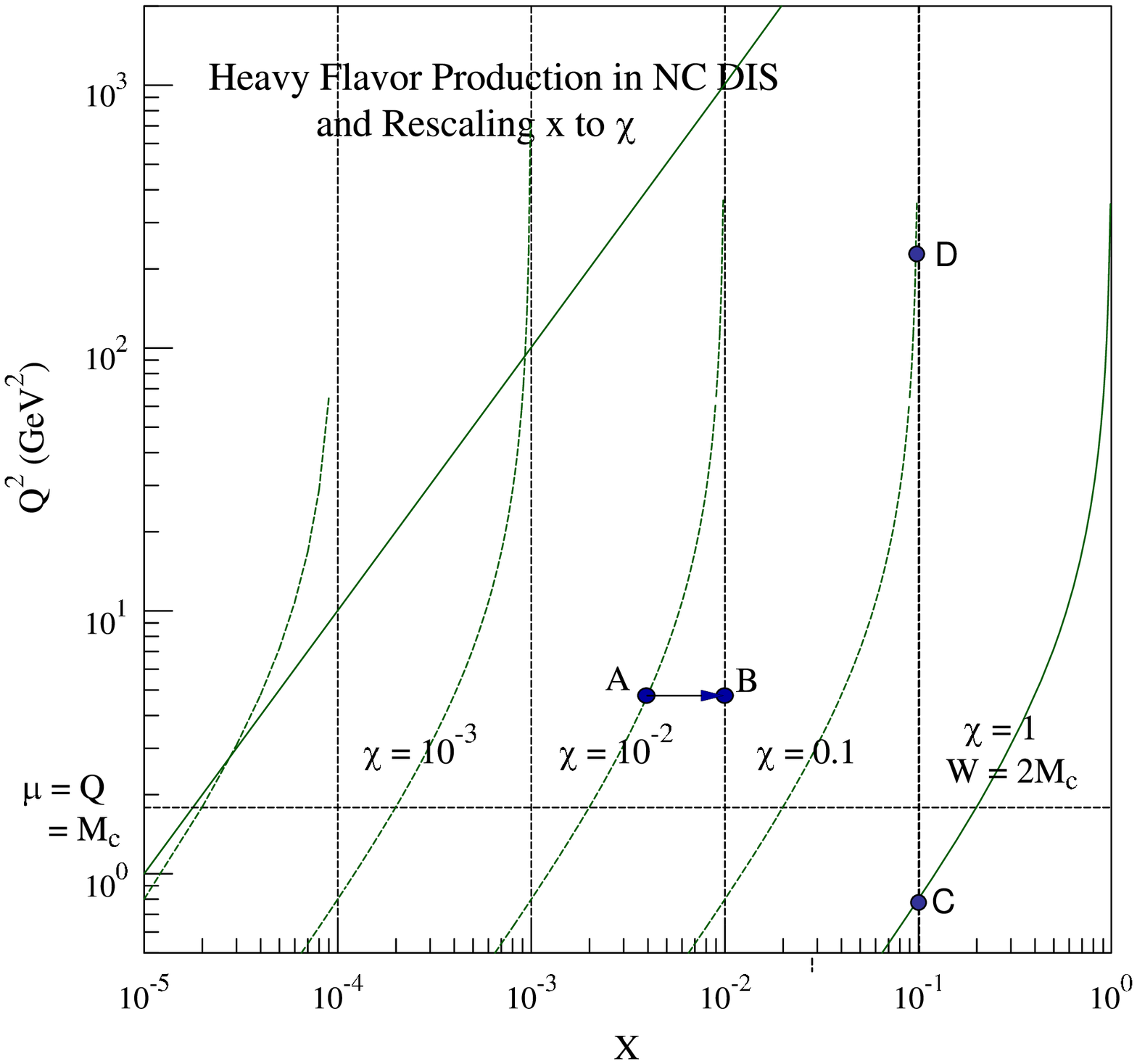}}
\caption{Kinematics of \emph{rescaling} due to $M_c$ in the $(x,Q)$
plane for NC DIS. The curves are constant $\chi$ ($\chi_c$) lines.
They asymptotically approach the corresponding Bjorken $x$ values
(vertical lines). The solid diagonal straight line marks the
kinematic reach of HERA.
  }%
 \label{fig:Rescaling}
}}
\newcommand{\figNloDiagm}
{
\FIGURE[h]{
 \centerline{\resizebox*{!}{0.35\textheight}{
 \includegraphics{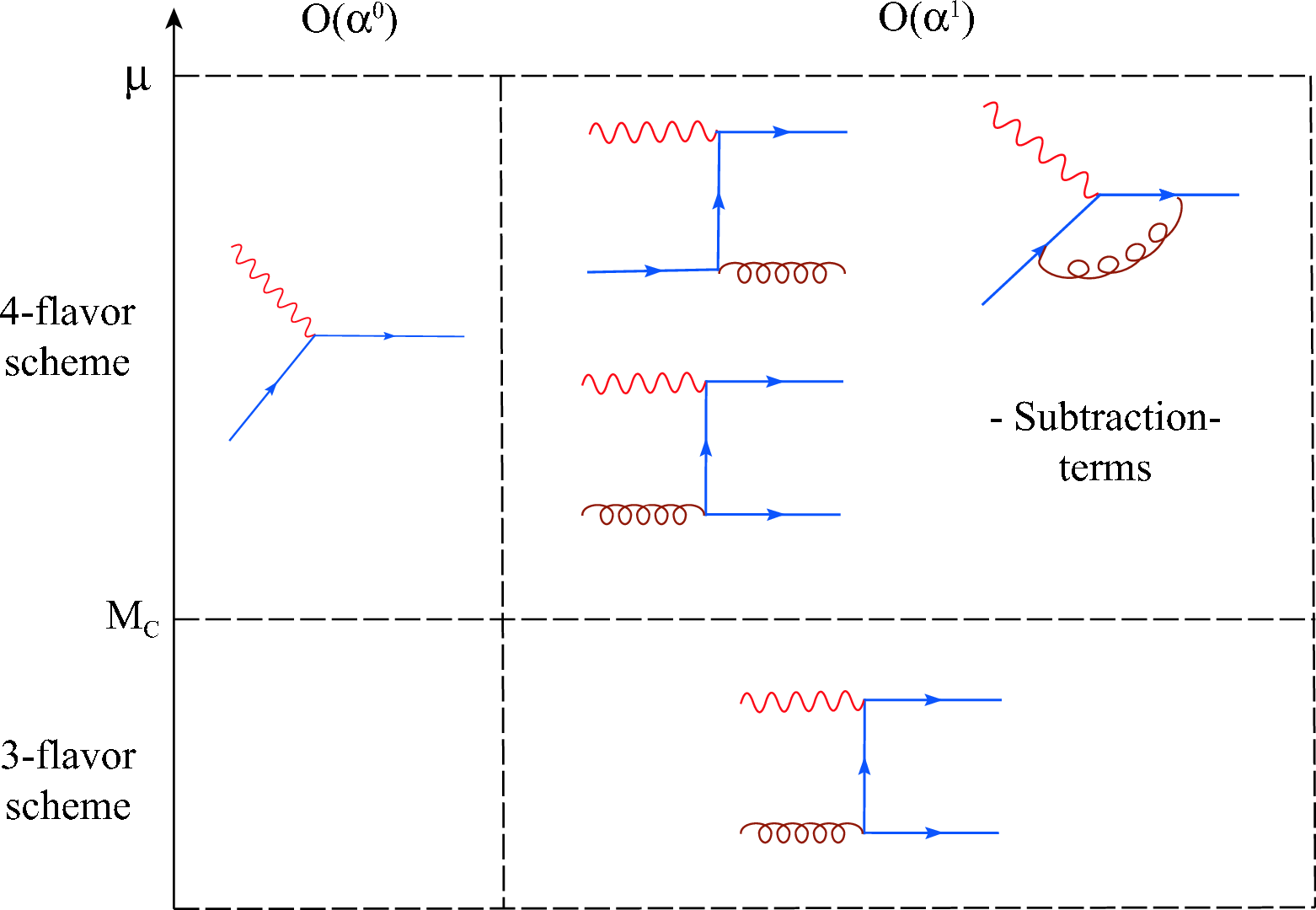}}
 }
\caption{Partonic processes included in the theoretical $n_f$-flavor
scheme calculations ($n_f=3,4$) at order $\alpha_s^0$ and
$\alpha_s^1$. Generalization to higher orders is straightforward.}
\label{fig:NloDiagm}
}
}
\newcommand{\figGmZmDiff}
{
\FIGURE[h]{
\hfill \resizebox*{!}{0.32\textheight}{\includegraphics{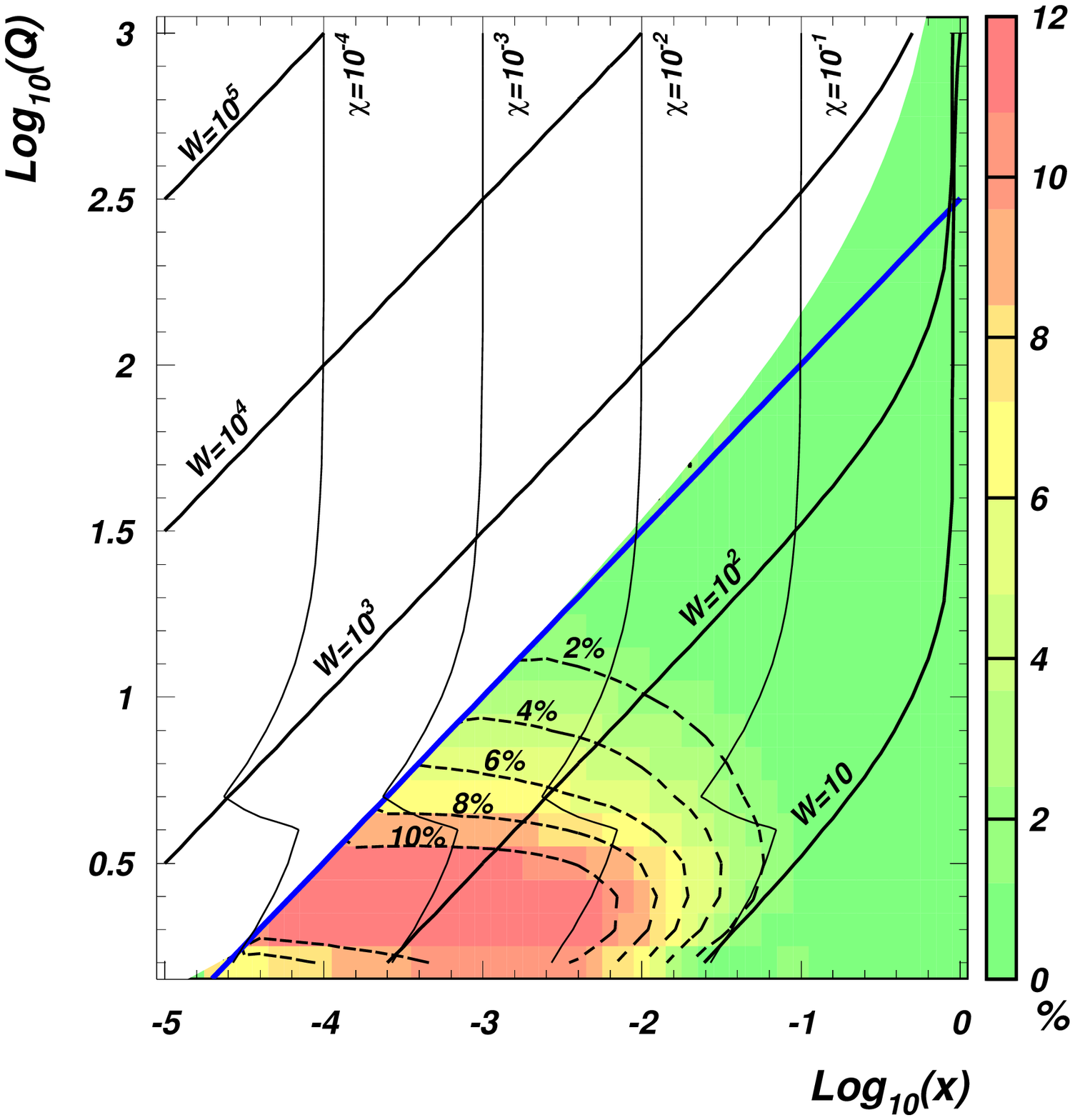}}
\hfill \resizebox*{!}{0.32\textheight}{\includegraphics{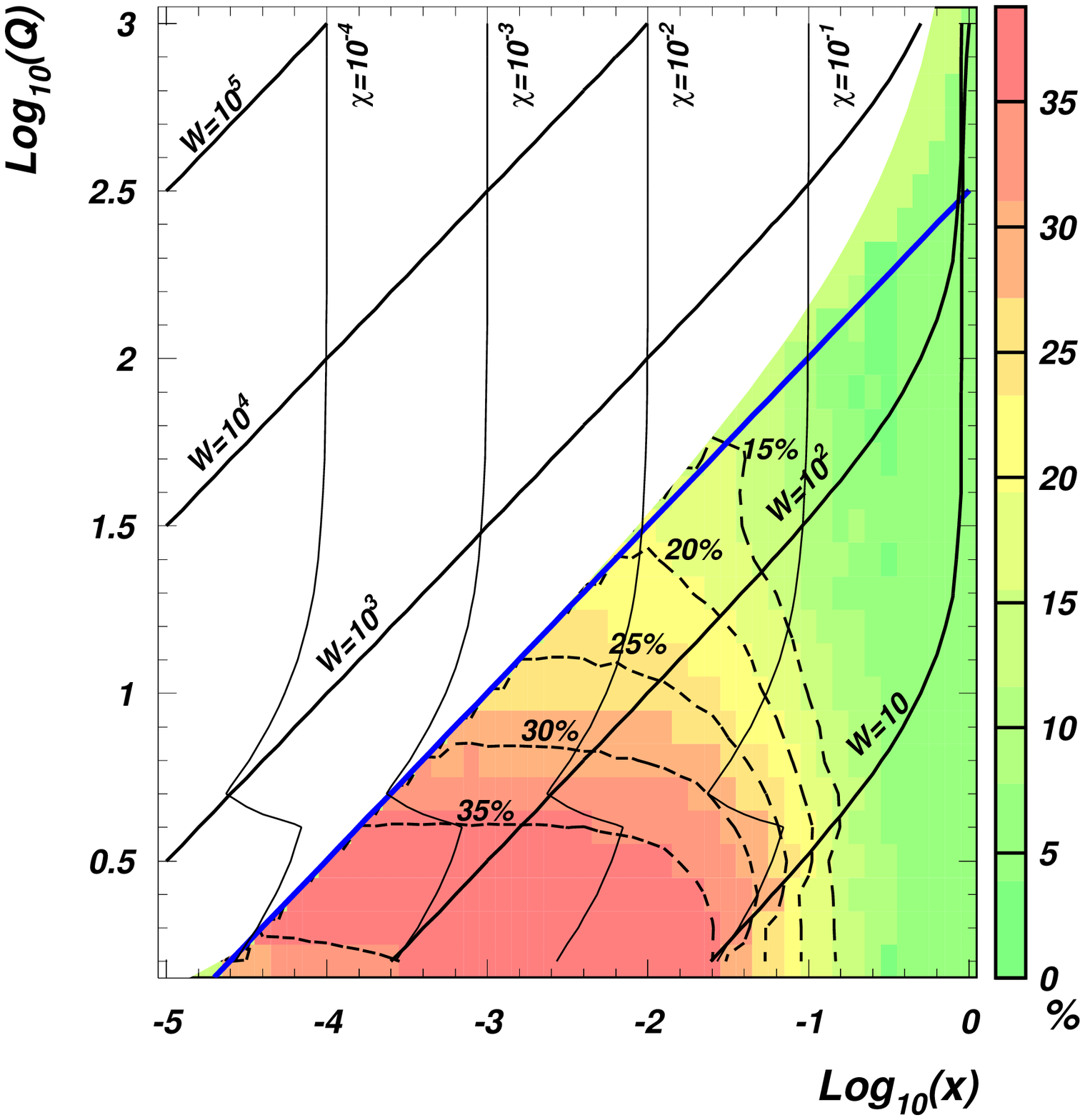}}
\hfill\rule{0ex}{1em} \caption{Fractional differences in percentages between
the general-mass (GM) and zero-mass (ZM) calculations of $F_2^\gamma$ (left
plot) and $F_L^\gamma$ (right plot), represented by color coding marked along
the right-side vertical axis and by the dashed contour curves.}
\label{fig:GmZmDiff}%
}
}
\newcommand{\figHoneTh}
{
\FIGURE[h]{
\resizebox*{0.48\textwidth}{!}{\includegraphics{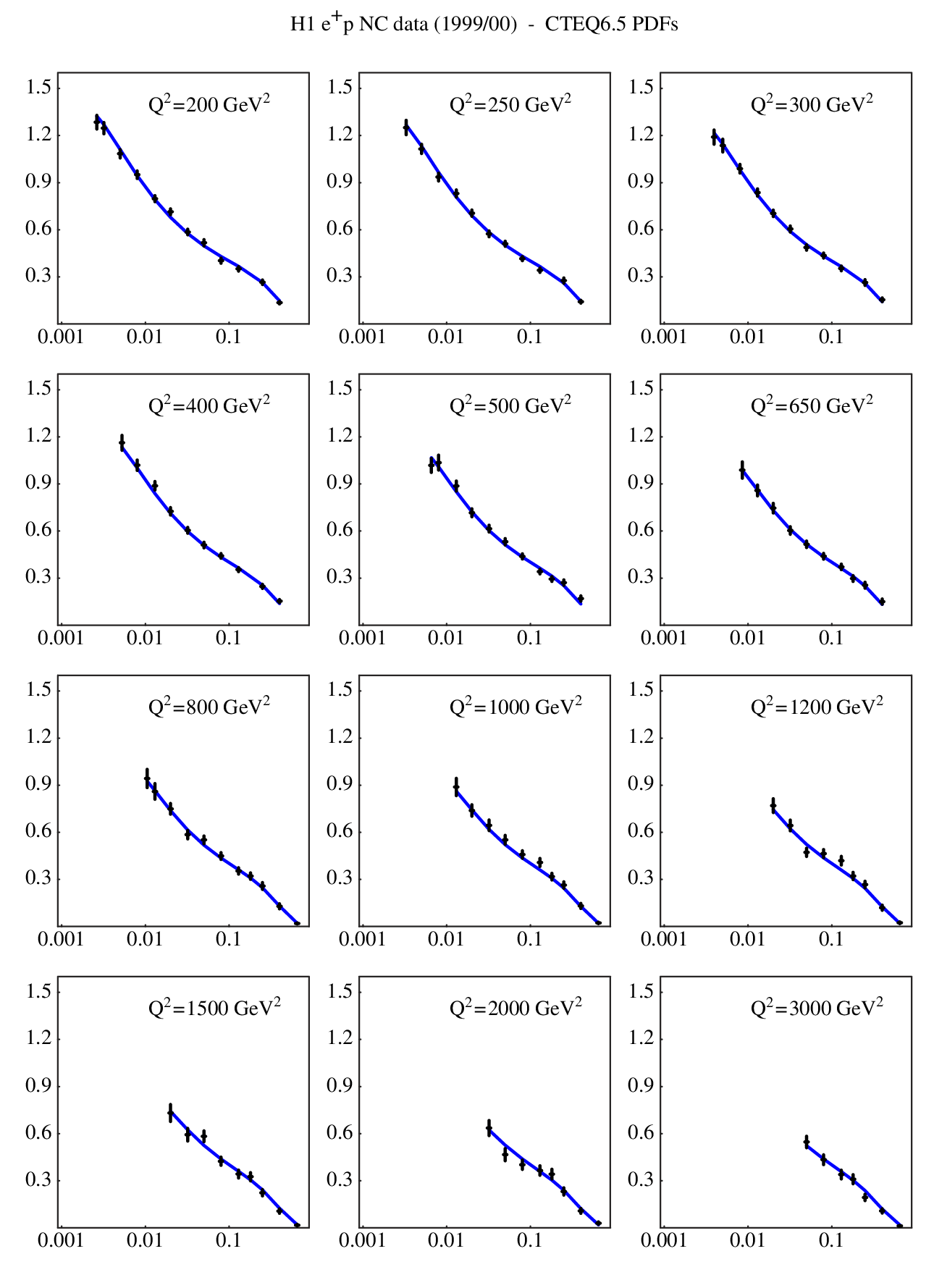}}
\hfill
\resizebox*{0.48\textwidth}{!}{\includegraphics{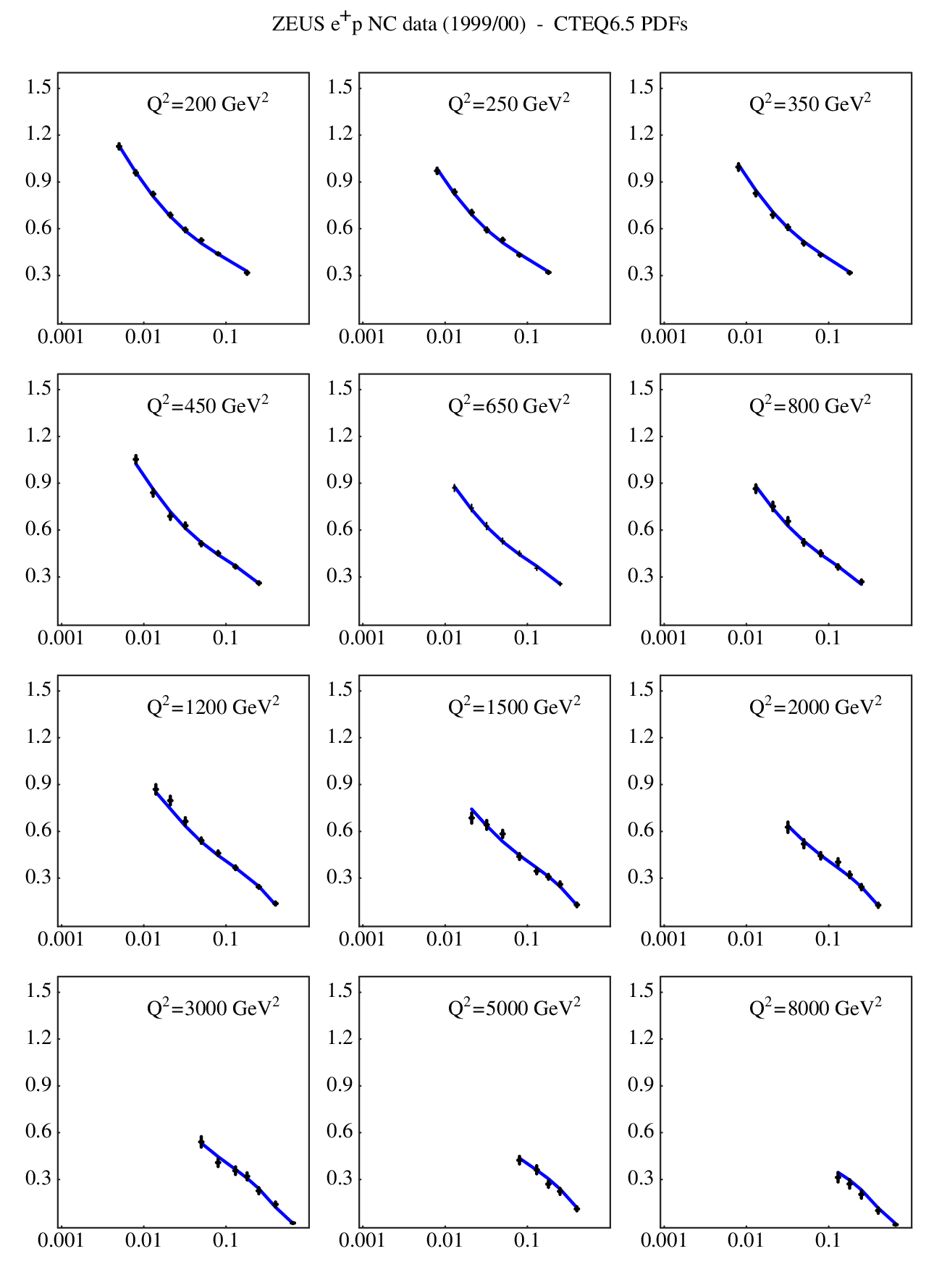}}
\caption{Comparison of the H1 (left) and ZEUS (right) 1999-2000 $e^{+}p$ NC
reduced cross section data sets with the CTEQ6.5M fit. } \label{fig:ZeusTh}
}
}
\newcommand{\figPdfNewOld}
{
\FIGURE[b]{
 \resizebox*{0.32\textwidth}{!}{
 \includegraphics{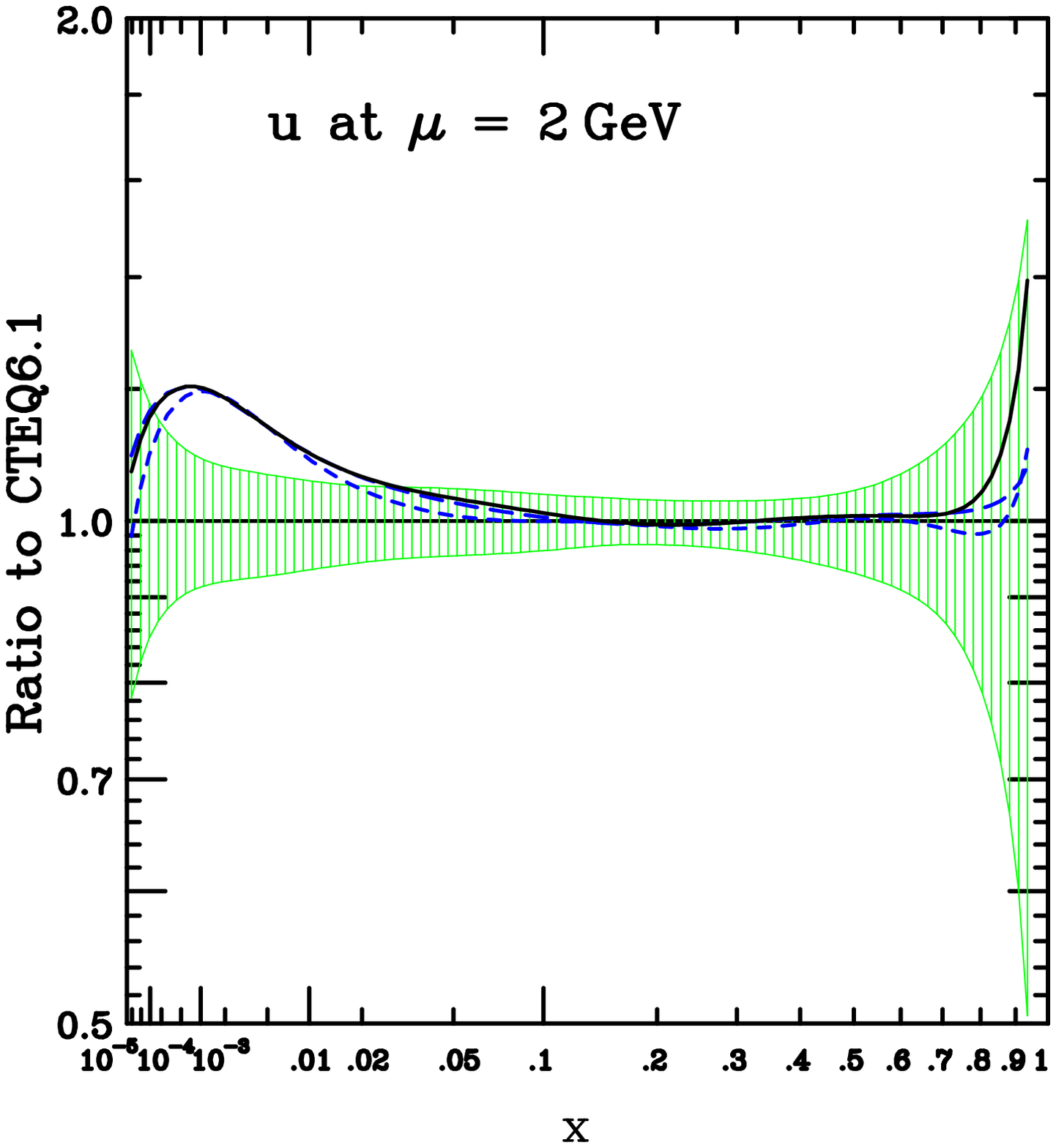}}\hfill
 \resizebox*{0.32\textwidth}{!}{
 \includegraphics{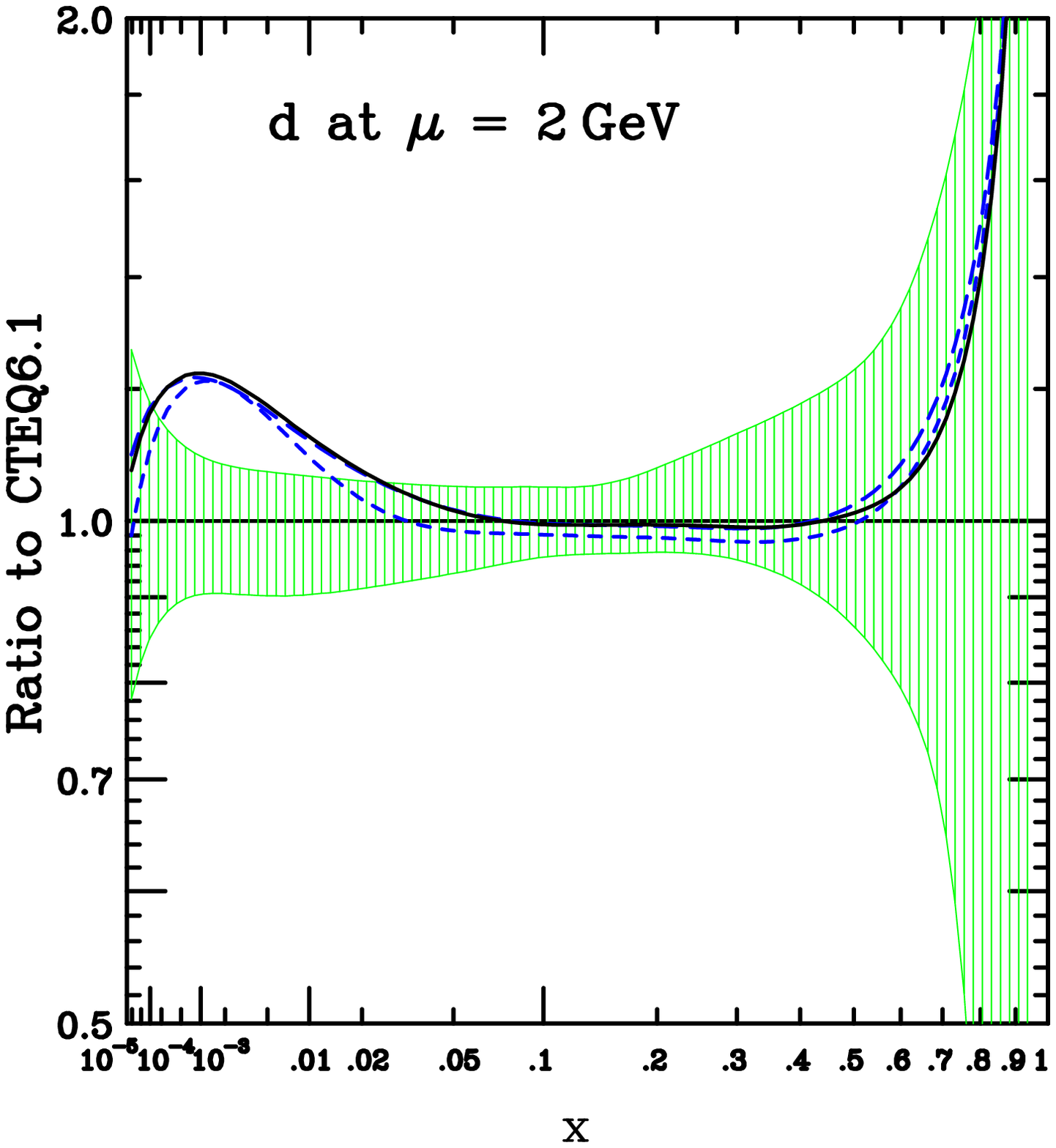}}\hfill
 \resizebox*{0.32\textwidth}{!}{
 \includegraphics{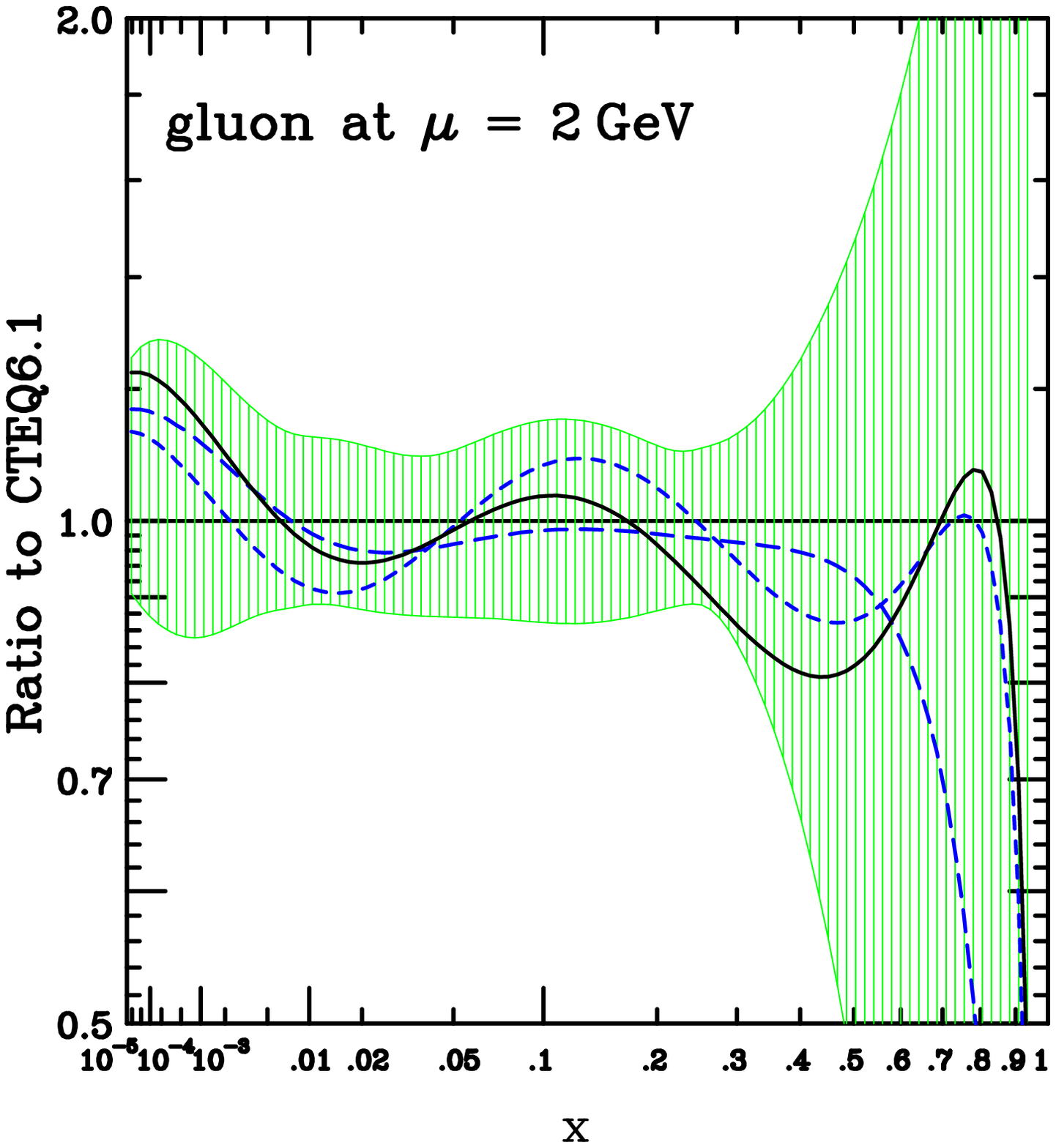}}
\caption{CTEQ6.5M $u,d$ and $g$ distributions (solid curves) at
scale $\mu=2$ GeV normalized to the corresponding ones from
CTEQ6.1M. The shaded areas represent the estimated uncertainty band
from the CTEQ6.1 analysis. The dashed curves represent alternative,
equally viable, candidates for this round of global analysis with
slightly different parametrization forms than CTEQ6.5M.}%
\label{fig:PdfNewOld}
}
}
\newcommand{\figQfive}
{
\FIGURE[hb]
{
 \resizebox*{0.32\textwidth}{!}{
 \includegraphics{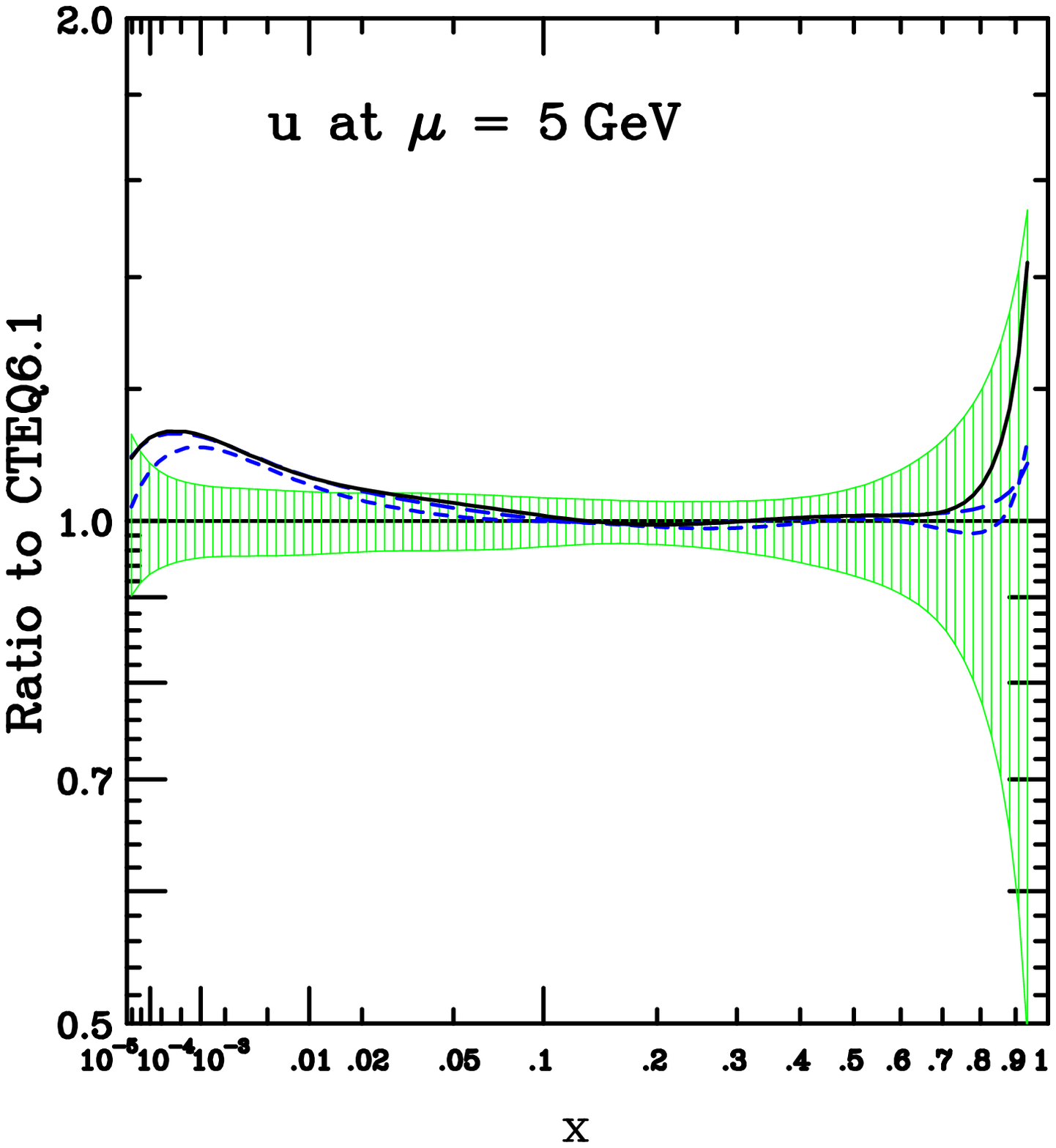}}\hfill
 \resizebox*{0.32\textwidth}{!}{
 \includegraphics{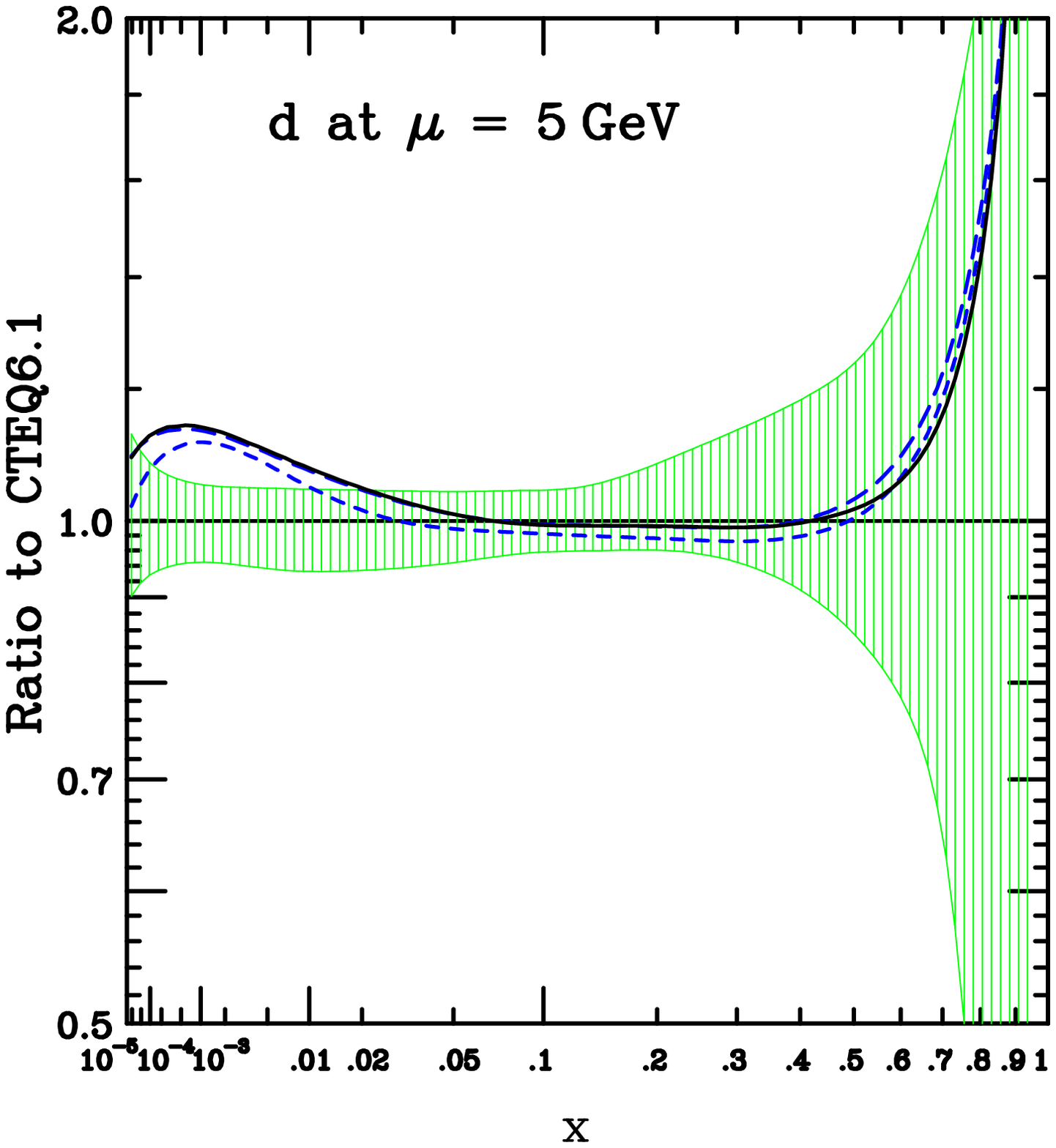}}\hfill
 \resizebox*{0.32\textwidth}{!}{
 \includegraphics{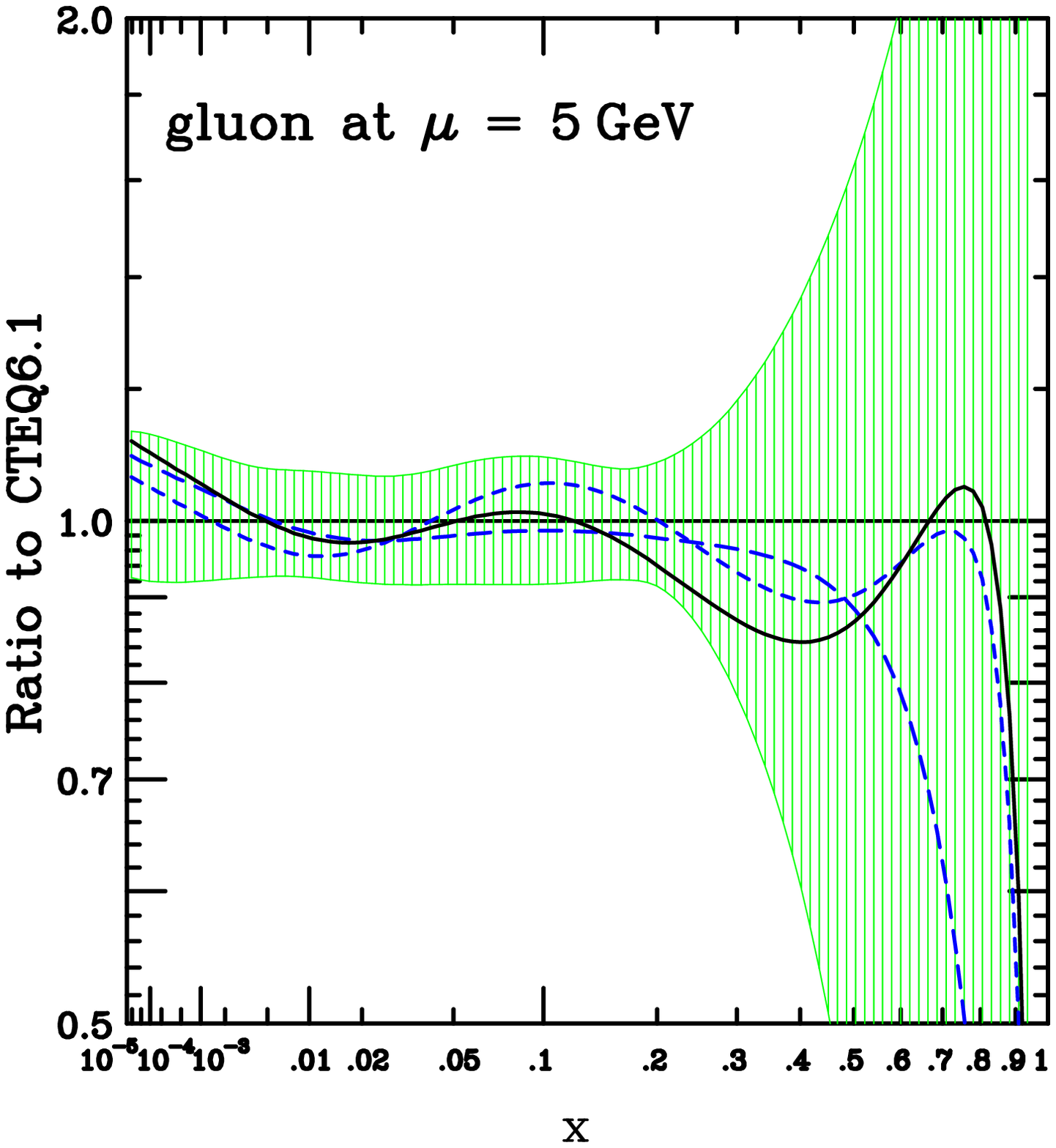}}
\caption{Same as Fig.\,\protect\ref{fig:PdfNewOld}, except that the PDFs are
at the scale $\mu=5$ GeV.} %
\label{fig:Qfive}
}
}
\newcommand{\figFtwocomQ}
{
\FIGURE[h]{
\resizebox*{0.6\textwidth}{!}{
 \includegraphics{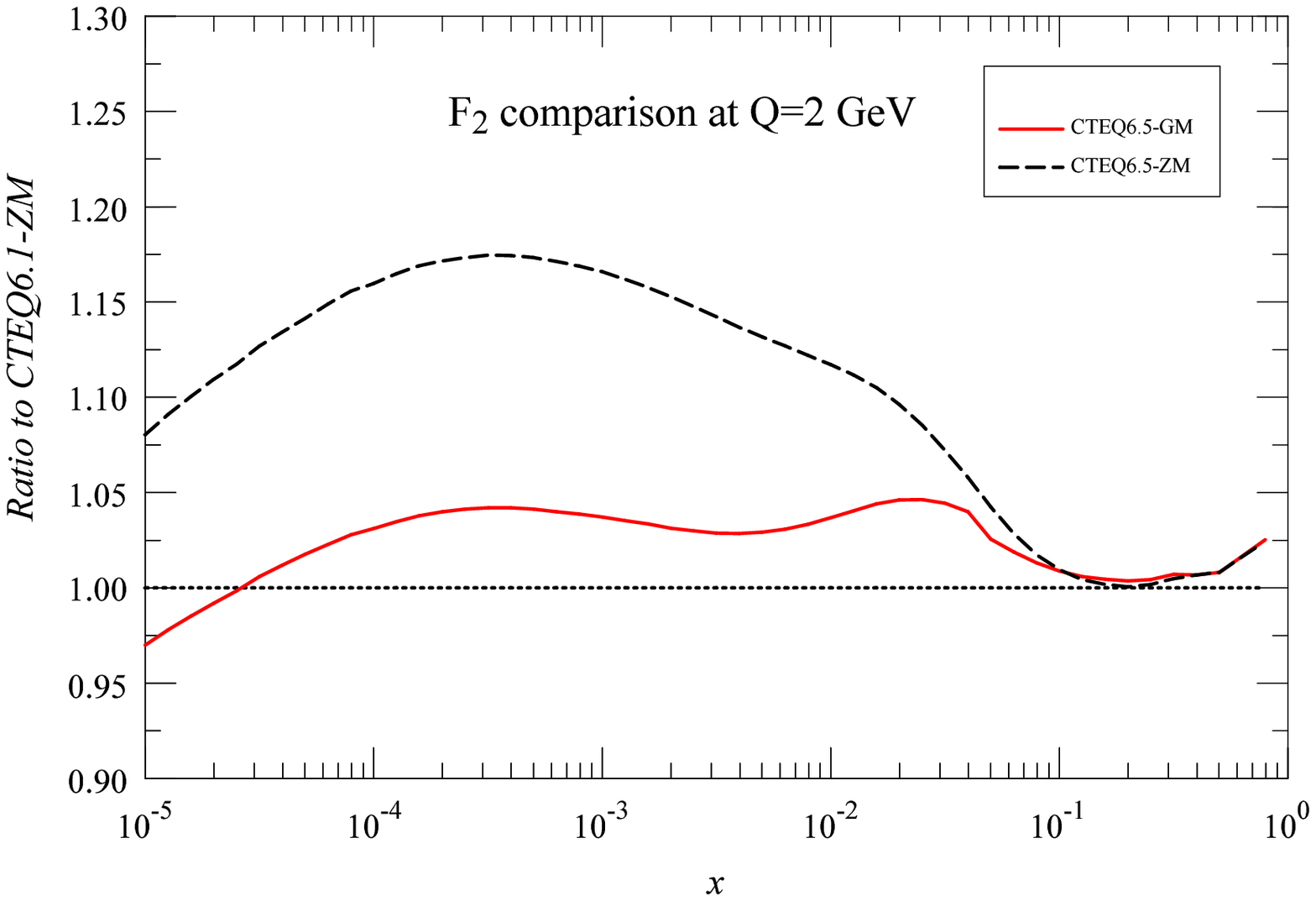}}
\caption{Comparison of theoretical calculations of $F_2$ using
CTEQ6.1M in the ZM formalism (horizontal line of 1.00), CTEQ6.5M in
the GM formalism (solid curve), and CTEQ6.5M in the ZM formalism
(dashed curve).} \label{fig:F2comQ}
}
}
\newcommand{\figLowQHera}
{
\FIGURE[h]{
\resizebox*{0.49\textwidth}{!}{
 \includegraphics{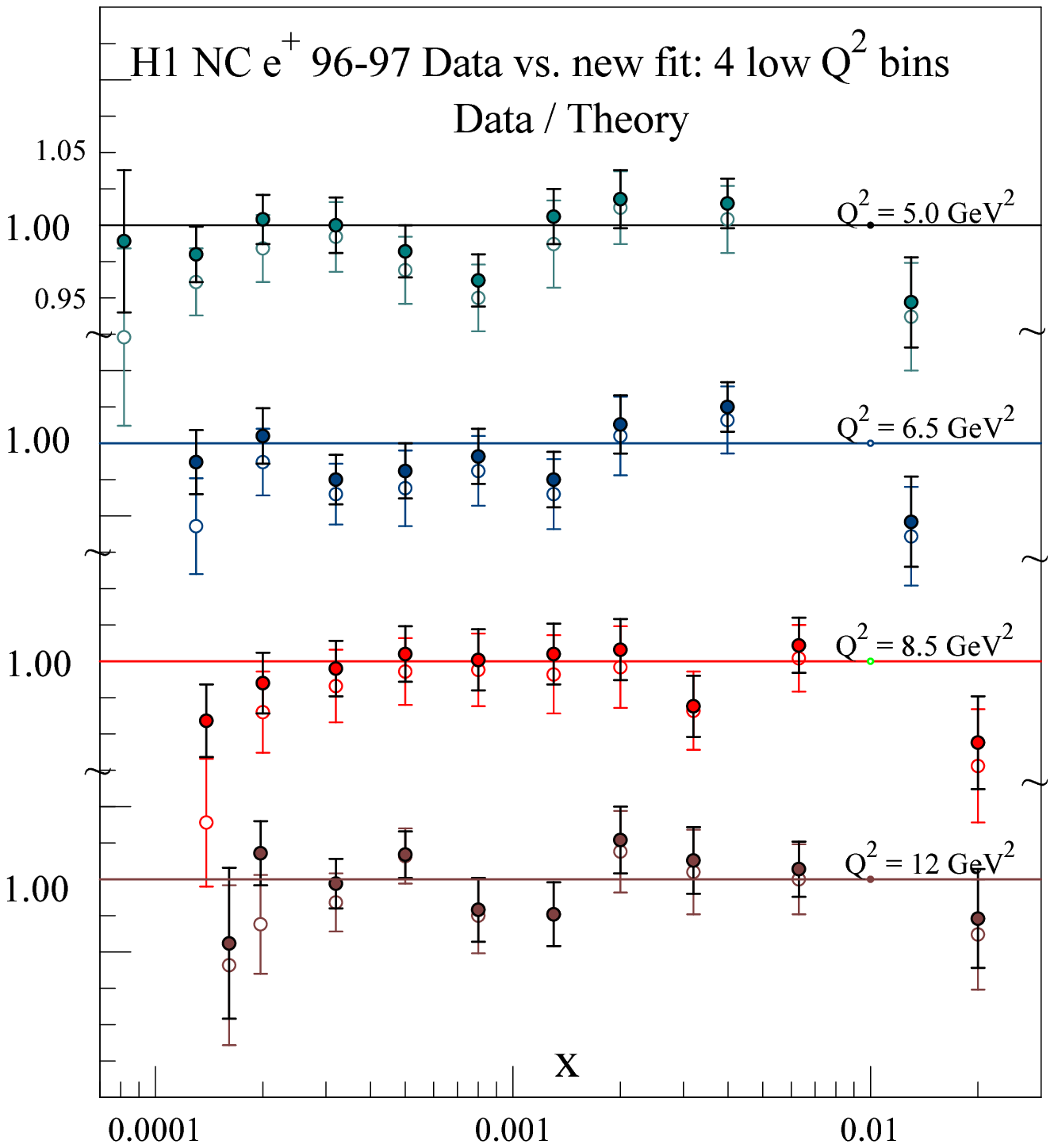}}
\hfill
\resizebox*{0.49\textwidth}{!}{
 \includegraphics{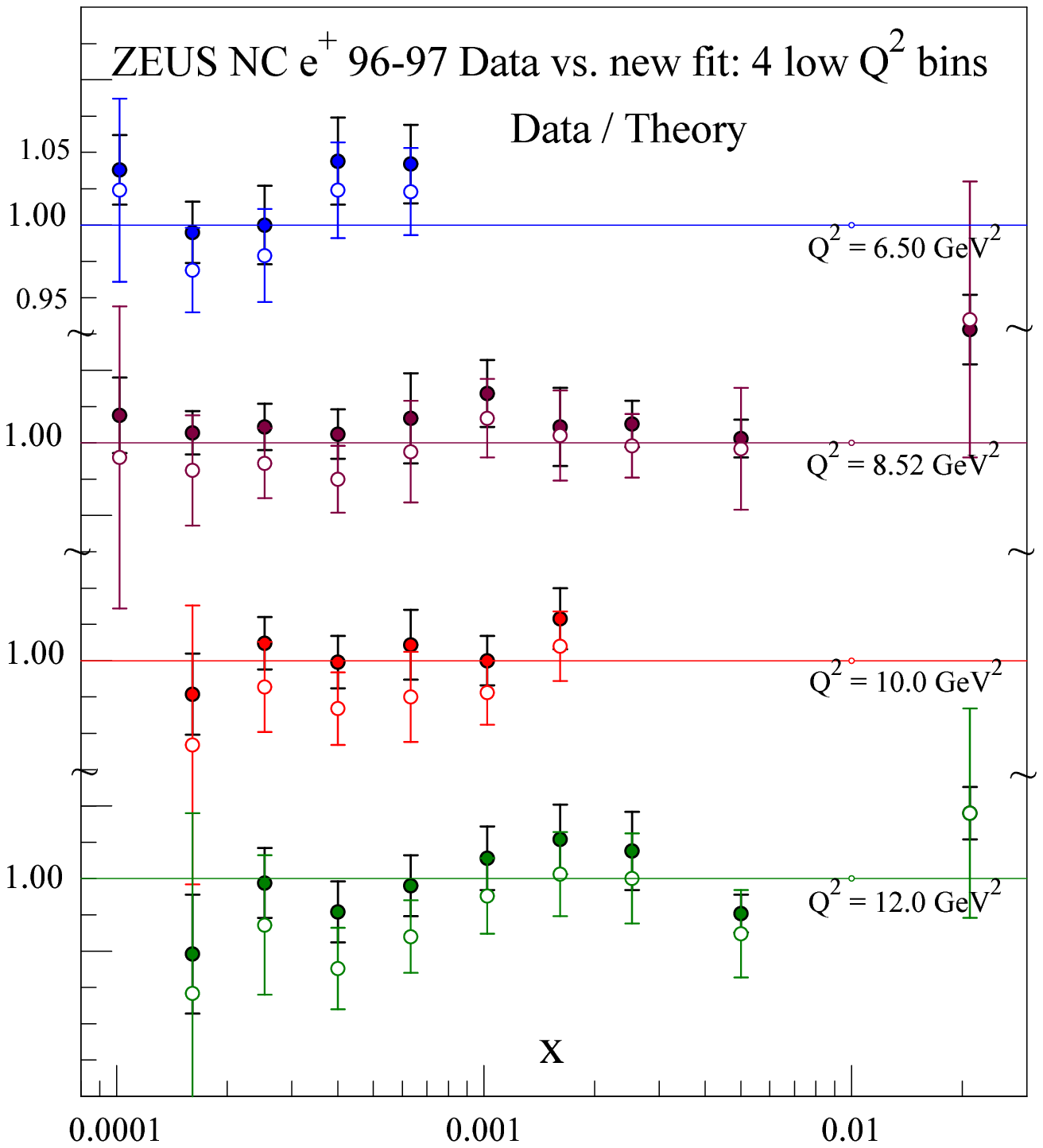}}
 \caption{Four low $Q$ bins of the H1 and ZEUS 1996-97 NC cross section
 data sets compared to the CTEQ6.5M fit. Open circles: the ratio data/theory; solid dots: the same data
 points shifted by correlated systematic errors. (See text.)} \label{fig:LowQHera}
}
}
\newcommand{\figRescale}
{
\FIGURE[h]{
 \centerline{\resizebox*{0.9\textwidth}{!}{
 \includegraphics{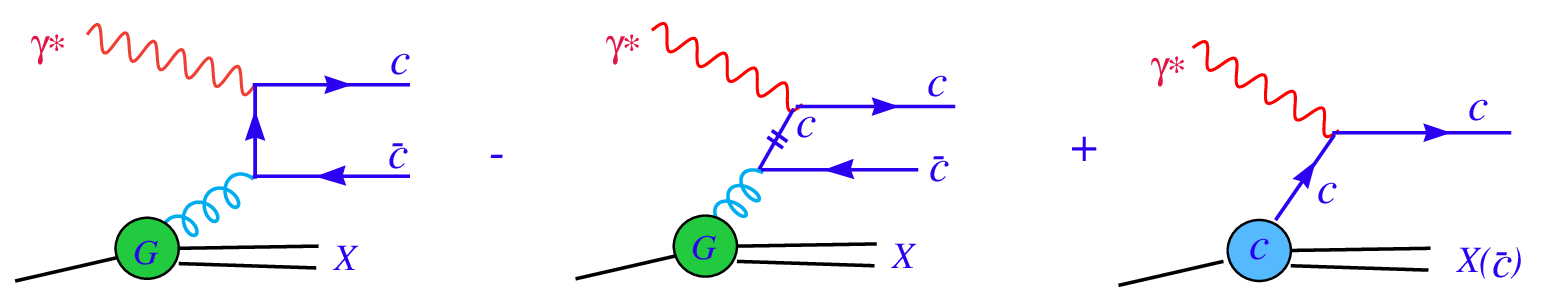}}
 }
\caption{Contributions to charm production in NC DIS scattering and
the physical origin of the rescaling variable in the PQCD approach.}
\label{fig:rescale}%
}
}
\newcommand{\figPdfBandA}
{
\FIGURE[h]{
 \resizebox*{0.32\textwidth}{!}{
 \includegraphics{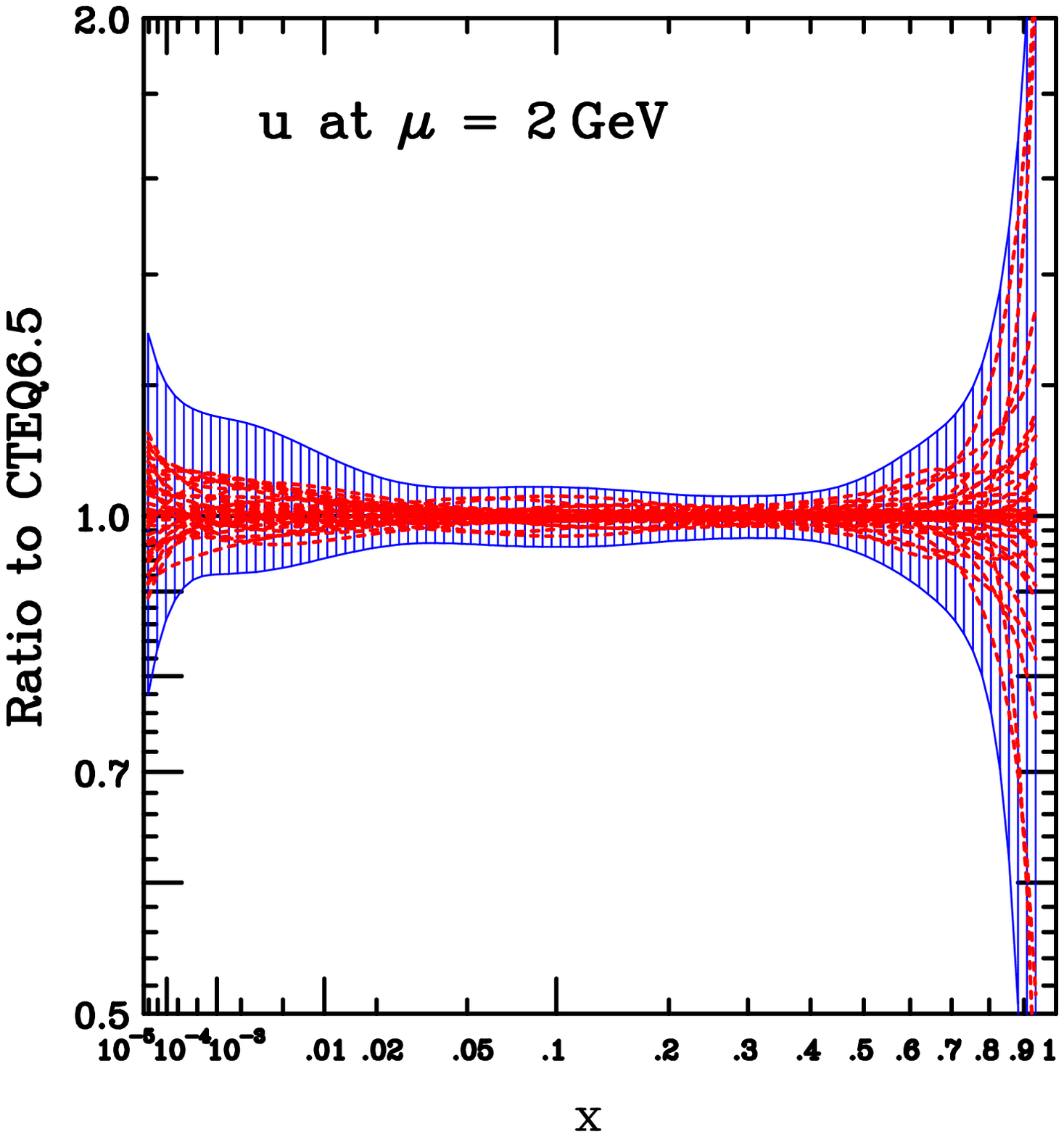}}\hfill
 \resizebox*{0.32\textwidth}{!}{
 \includegraphics{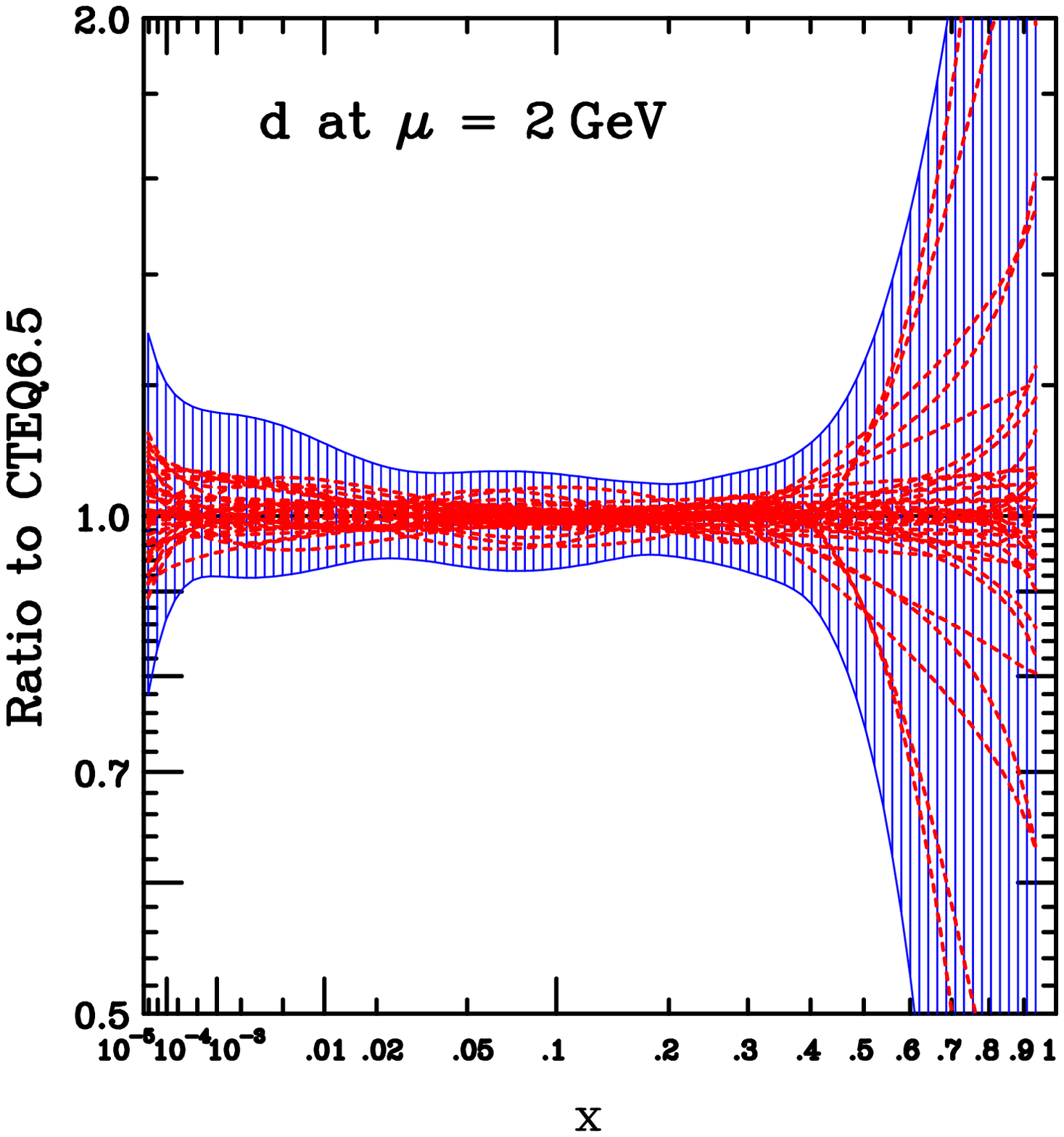}}\hfill
 \resizebox*{0.32\textwidth}{!}{
 \includegraphics{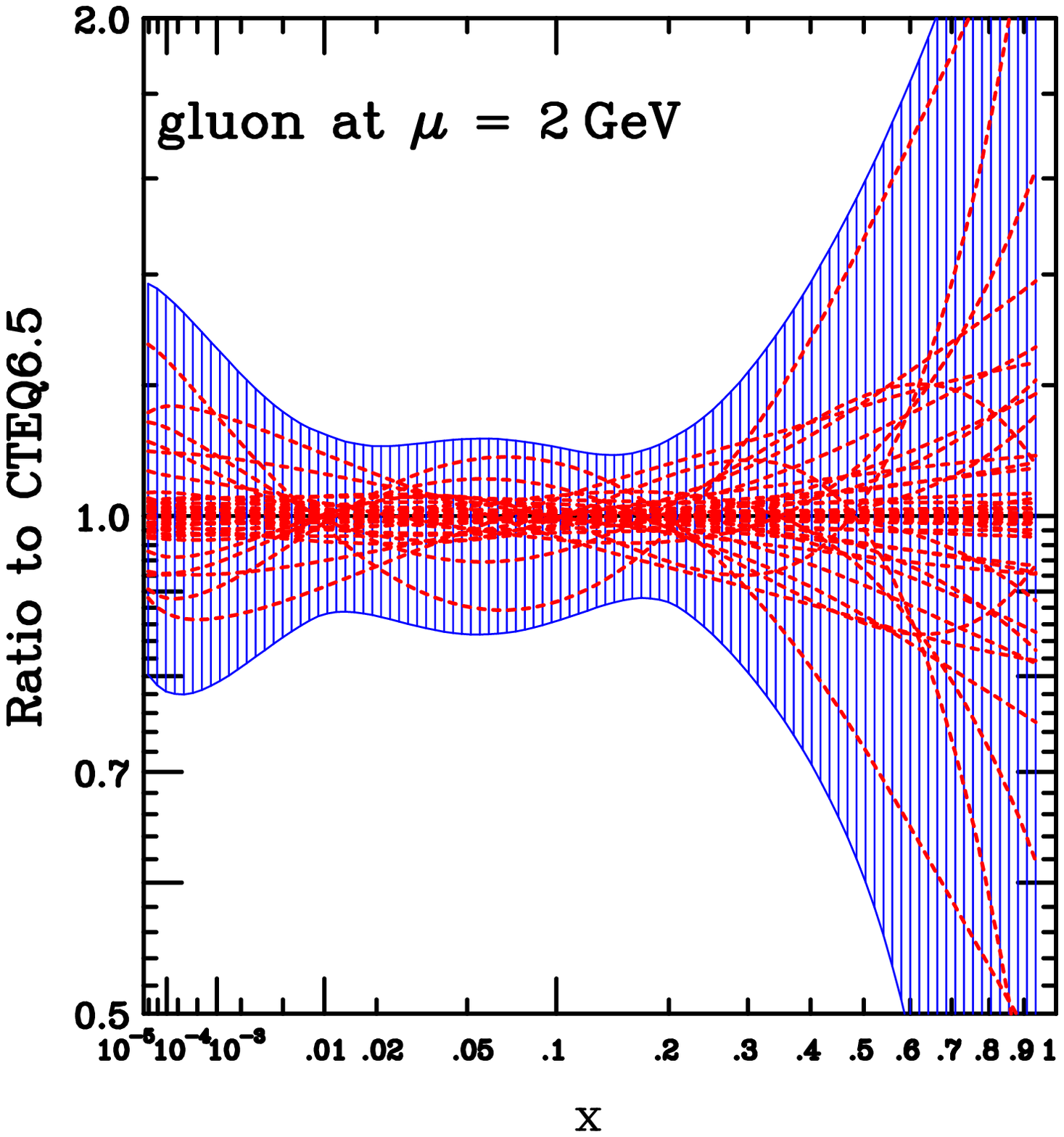}}
\caption{CTEQ6.5 PDF uncertainty bands and eigenvector PDF sets.} \label{fig:PdfBandA}
}
}
\newcommand{\figPdfBandB}
{
\FIGURE[h]{
 \resizebox*{0.32\textwidth}{!}{
 \includegraphics{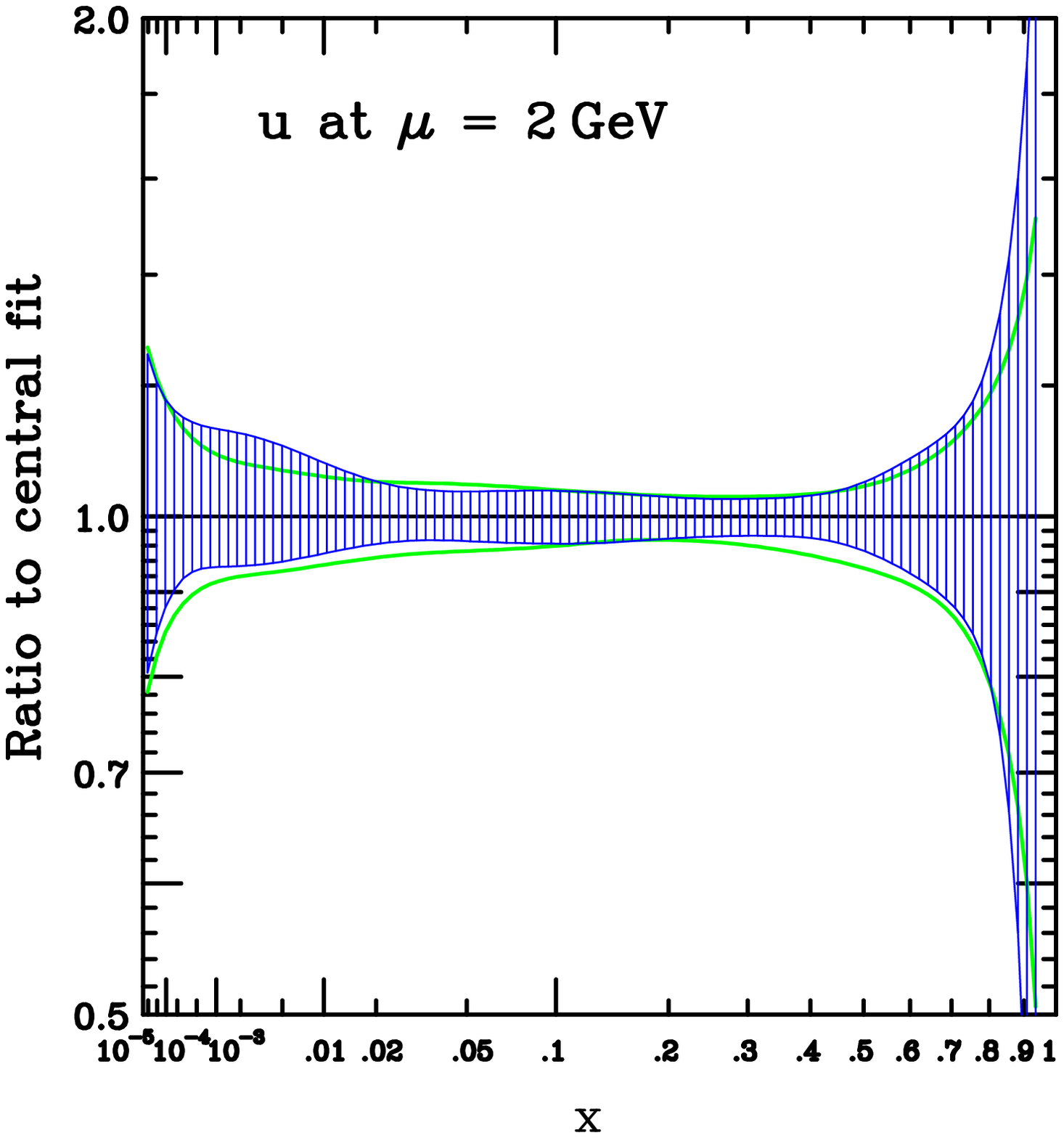}}\hfill
 \resizebox*{0.32\textwidth}{!}{
 \includegraphics{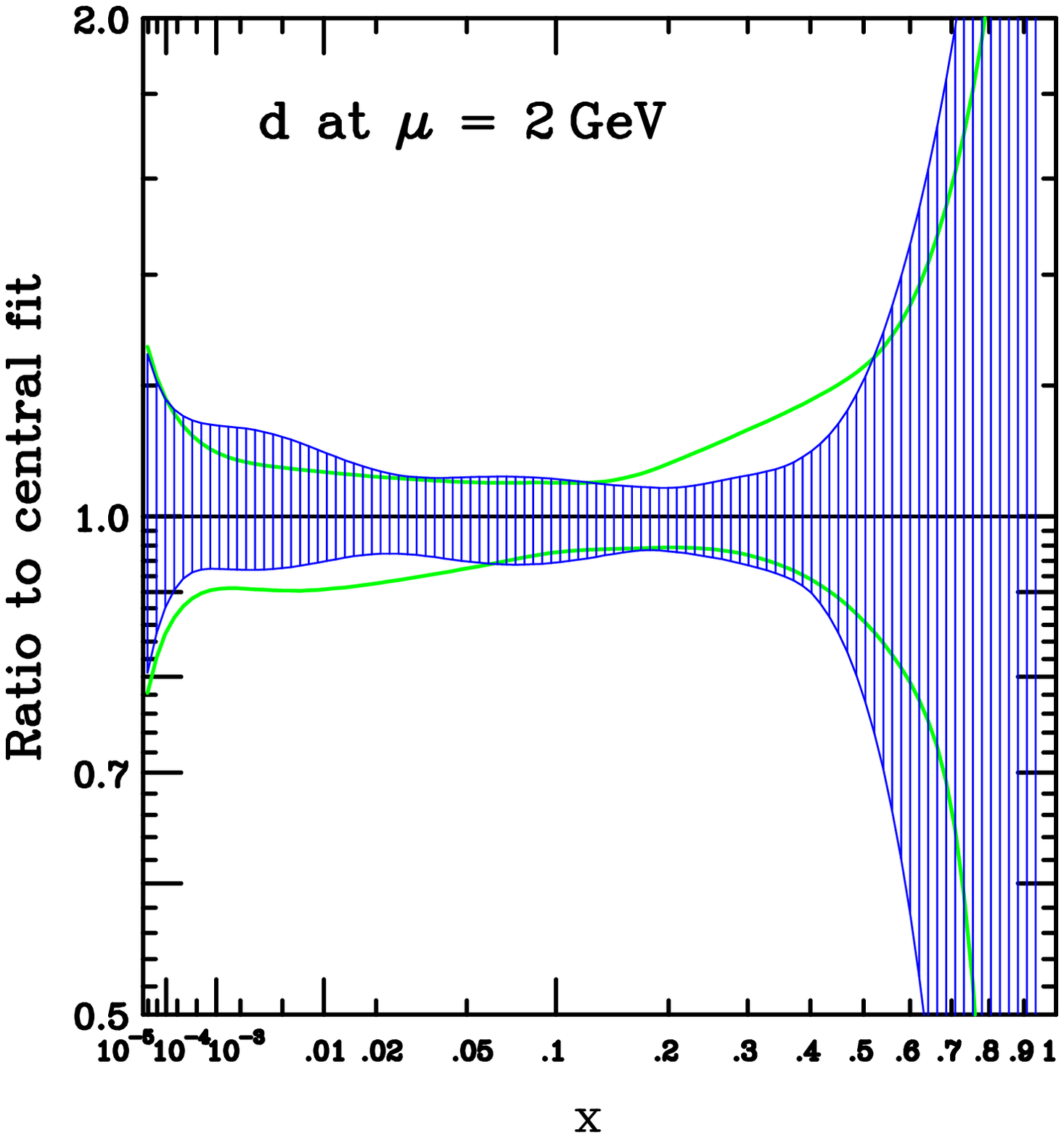}}\hfill
 \resizebox*{0.32\textwidth}{!}{
 \includegraphics{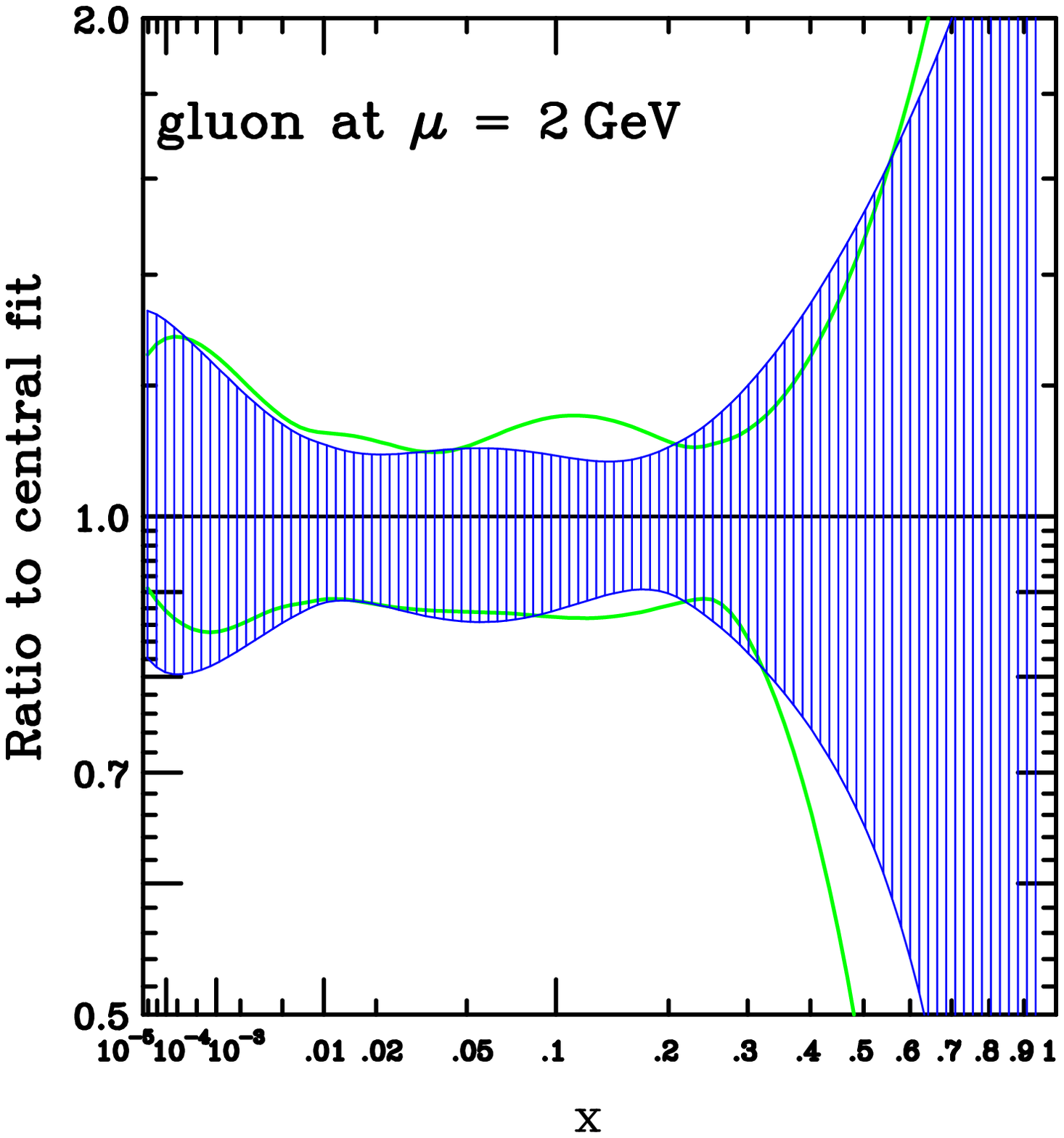}}
\caption{CTEQ6.5 PDF uncertainty bands compared to those of CTEQ6.1.} \label{fig:PdfBandB}
}
}
\newcommand{\figPdfBandC}
{
\FIGURE[h]{
 \resizebox*{0.32\textwidth}{!}{
 \includegraphics{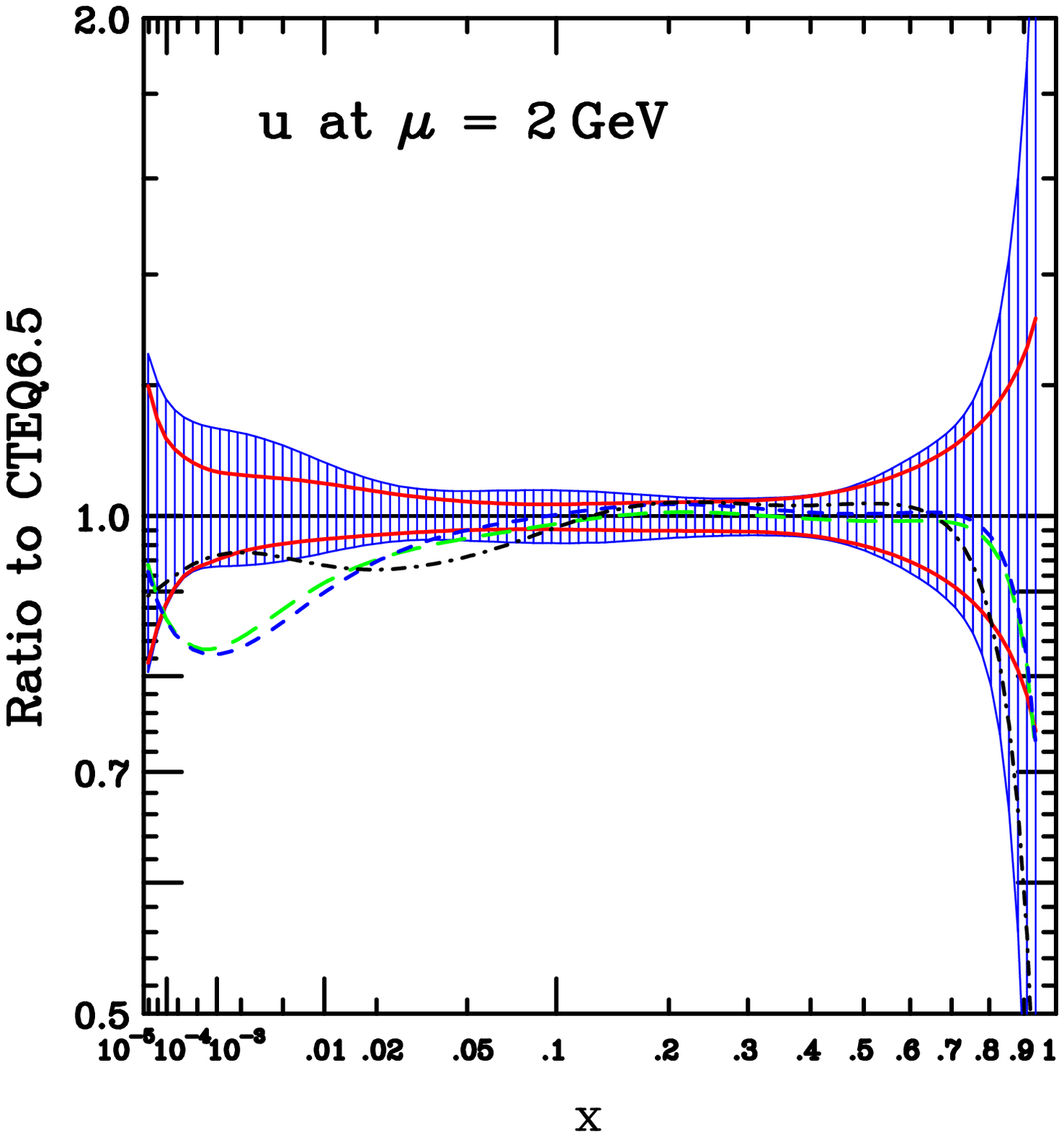}}\hfill
 \resizebox*{0.32\textwidth}{!}{
 \includegraphics{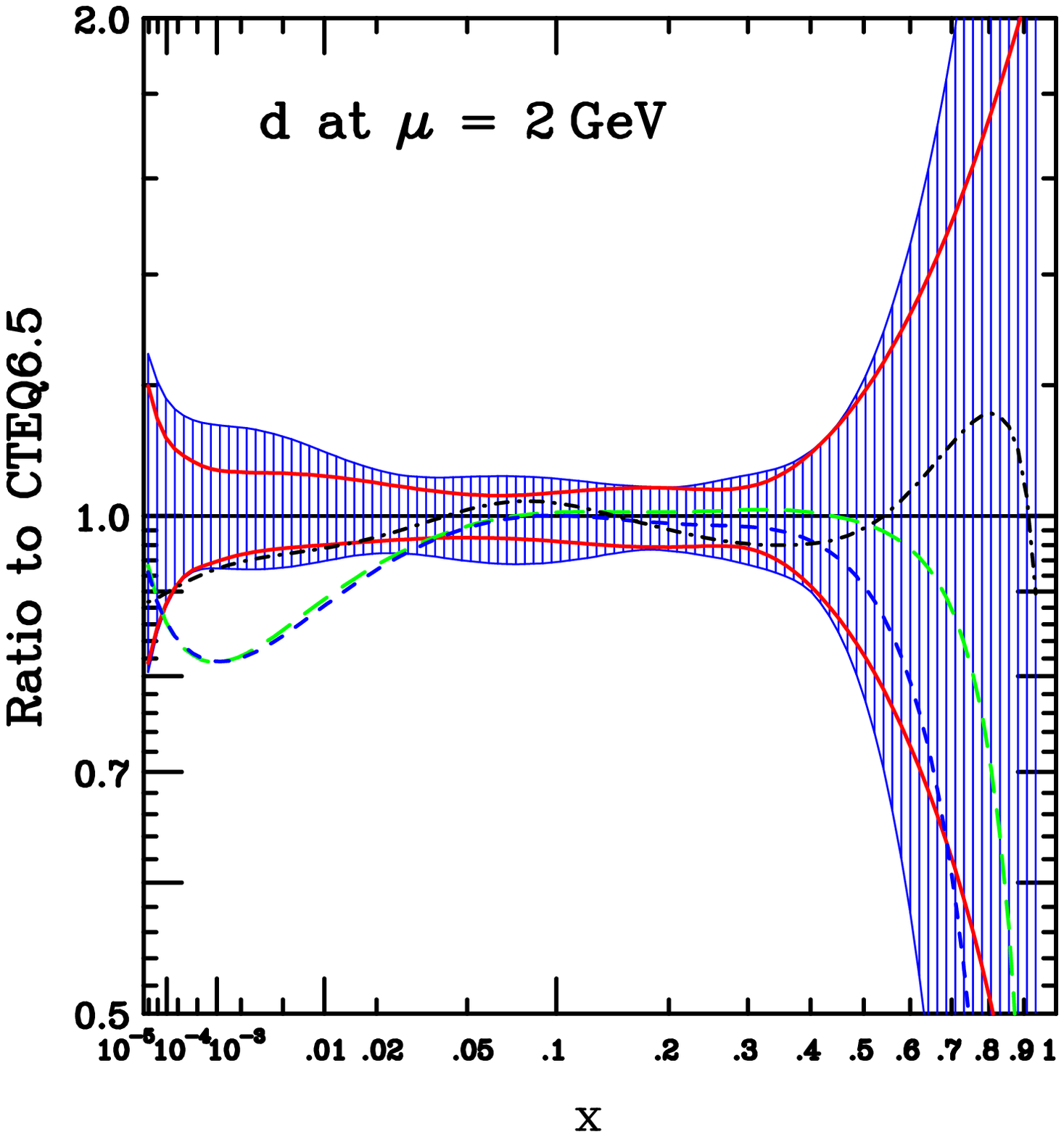}}\hfill
 \resizebox*{0.32\textwidth}{!}{
 \includegraphics{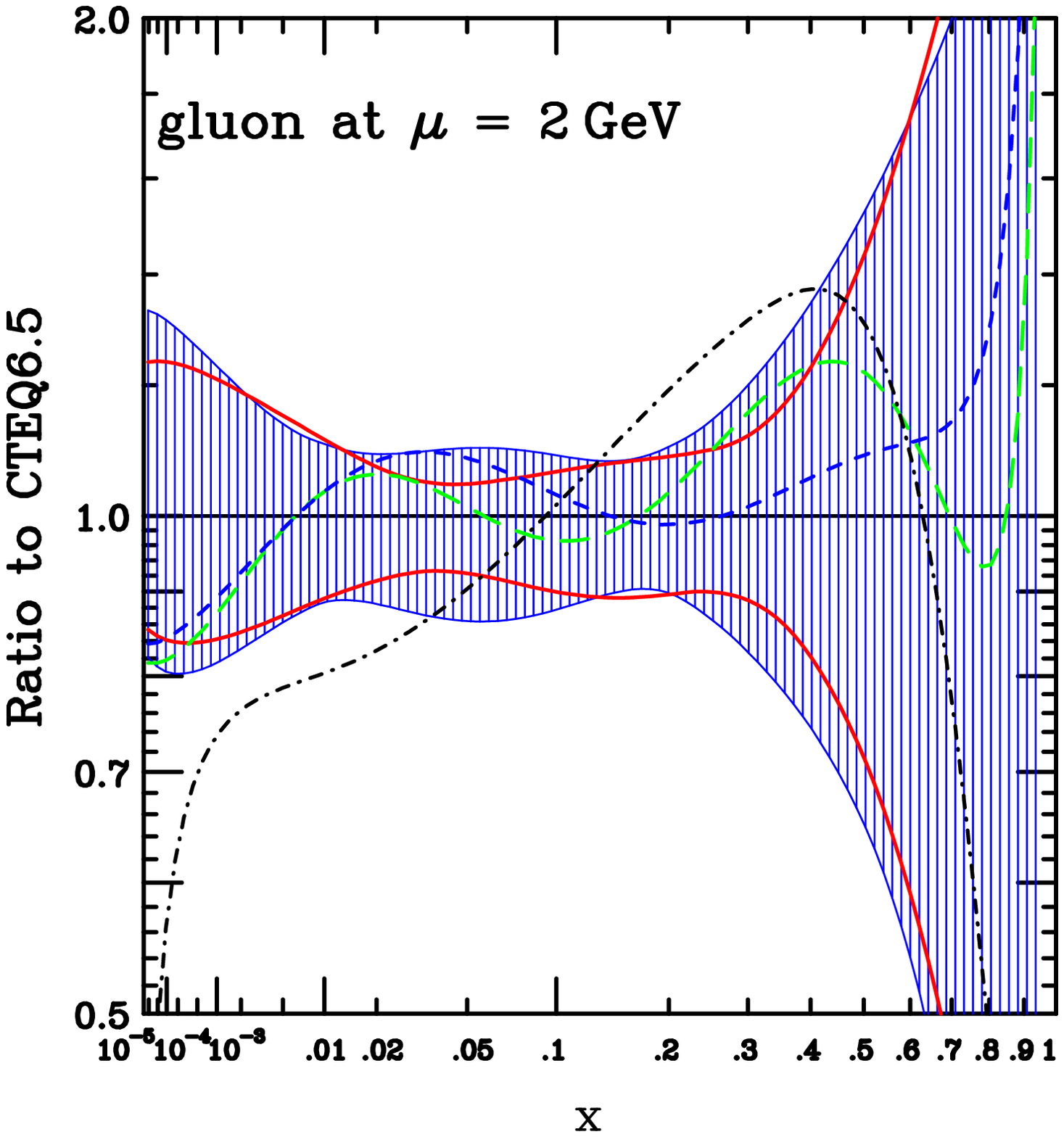}}
\caption{Same CTEQ6.5 PDF uncertainty bands compared to: (i) the
CTEQ6.1M (green long-dash line), CTEQ6A118 (blue short-dash)
and MRST04 (black dash-dotted) PDFs; and (ii)
upper and lower edges (red solid lines) of similar uncertainty bands
generated with reduced number of fitting parameters (see text).}
\label{fig:PdfBandC}
}
}
\newcommand{\tblHoneData}
{
\TABLE[bh]{
\rule{0em}{1ex}\hfill
\begin{tabular}{|c|c|c|c|c|}
\hline \multicolumn{5}{|c|}{H1} \\ \hline 94-97 &  & CC & $\sigma
_{tot}$ & \cite{H1N+67X0} \\ \hline 96-97 &  & NC & $\sigma _{tot}$
& \cite{Adloff:2000qk} \\ \hline &  & NC & $F_{2}^{c}$ &
\cite{H1N+67F2c} \\ \hline 98-99 & $e^{-}p$ & NC & $\sigma _{tot}$ &
\cite{Adloff:2000qj} \\ \hline &  & CC & $\sigma _{tot}$ &
\cite{Adloff:2000qj} \\ \hline 99-00 & $e^{+}p$ & NC & $\sigma
_{tot}$ & \cite{Adloff:2003uh} \\ \hline &  & CC & $\sigma _{tot}$ &
\cite{Adloff:2003uh} \\ \hline &  & NC & $\sigma _{c}$ &
\cite{H1N+90X0bcHQ,H1N+90X0bcLQ} \\ \hline &  & NC & $\sigma _{b}$ &
\cite{H1N+90X0bcHQ,H1N+90X0bcLQ} \\ \hline & $e^{-}p$ & NC & $\sigma
_{tot}$ & \cite{Adloff:2003uh} \\ \hline
\end{tabular}%
\hfill
\begin{tabular}{|c|c|c|c|c|}
\hline
\multicolumn{5}{|c|}{ZEUS} \\ \hline
94-97 & $e^{+}p$ & CC & $\sigma _{tot}$ & \cite{Breitweg:1999aa} \\ \hline
96-97 &  & NC & $\sigma _{tot}$ & \cite{Chekanov:2001qu} \\ \hline
&  & NC & $F_{2}^{c}$ & \cite{ZN+67F2c} \\ \hline
98-99 & $e^{-}p$ & NC & $\sigma _{tot}$ & \cite{Chekanov:2002ej} \\ \hline
&  & CC & $\sigma _{tot}$ & \cite{Chekanov:2002zs} \\ \hline
&  & NC & $F_{2}^{c}$ & \cite{ZN+80F2c} \\ \hline
99-00 & $e^{+}p$ & NC & $\sigma _{tot}$ & \cite{Chekanov:2003yv} \\ \hline
&  & CC & $\sigma _{tot}$ & \cite{Chekanov:2003vw} \\ \hline
\end{tabular}
\hfill \rule{0em}{1ex}
\caption{HERA I table sets used in the global analysis.}
}}

%% file: text/1.intro.tex


\section{Introduction\label{sec:intro}}

Global QCD analysis of the parton structure of the nucleon has made
significant progress in recent years. However, there remain many gaps in our
knowledge of the parton distribution functions (PDFs), especially with regard
to the strange, charm and bottom degrees of freedom. The uncertainties of the
PDFs remain large in both the very small-$x$ and the large-$x$ regions---in
general for all flavors, but particularly for the gluon. \ Uncertainties due
to the input PDFs will be one of the dominant sources of uncertainty in some
precision measurements (such as the $W$ mass) \cite{Group:2006rt}, as well as
in studying signals and backgrounds for New Physics searches at the Fermilab
Tevatron and the CERN Large Hadron Collider (LHC). Thus, improving the
accuracy of the global QCD analysis of PDFs is a high priority task for High
Energy Physics.

With the accumulation of extensive precision deep inelastic scattering (DIS)
cross section measurements of both the neutral current (NC) and charged
current (CC) processes at HERA I (and even more precise data from HERA II to
come soon), it is necessary to employ reliable theoretical calculations that
match the accuracy of the best data in the global analysis. \ In the
perturbative QCD framework (PQCD), this requires, among other things, a
proper treatment of the heavy quark mass parameters.\footnote{ Our
discussions will be independent of the flavor of the heavy quark. In
practice, \textquotedblleft heavy quarks\textquotedblright\ means charm and
bottom. The top quark is so heavy that it can generally be treated as a heavy
particle, not a parton.} There are many aspects to a proper treatment of
general mass effects, involving both dynamics (consistent factorization in
PQCD, with quark massess) and kinematics (physical phase space constraints
with heavy flavor masses that are not satisfied by the simplest
implementation of the zero-mass parton model formula). Some aspects of these
considerations have been applied in existing work on global QCD analysis;
however, in most cases, not {\em all} relevant effects have been consistently
taken into account.

In this paper, we present a systematic discussion of the relevant
physics issues, and provide a new implementation of the general
formalism for all DIS processes in a unified framework
(Sec.\,\ref{sec:implement}). We show the magnitude of the mass
effects, compared to the conventional zero-mass (ZM) parton
formalism (Sec.\,\ref{sec:difference}). We then apply this simple
implementation of the general mass formalism to a precise global
analysis of PDFs, including the full HERA I cross section data sets
for both NC and CC processes, and taking into account all available
correlated systematic errors (Sec.\,\ref{sec:globalfit}). We
investigate in some depth the parametrization dependence of the
global analysis, assess the uncertainties of the PDFs in the new
analysis, and compare the new results with those of CTEQ6.1
\cite{Cteq6} (which was based on the ZM parton formalism) and other
current PDFs \cite{Mrst04PhyG} (Sec.\,\ref{sec:uncertainty}). The
results, designated as CTEQ6.5 PDFs, have important implications for
hadron collider phenomenology at the Tevatron and the LHC. Finally,
we summarize our main results, discuss their limitations, and
mention the challenges ahead (Sec.\,\ref{sec:summary}). Some
preliminary results of this investigation have been presented at the
DIS2006 Workshop \cite{Wkt06Dis1}.

%% file: text/2.implement.tex


\section{General Mass PQCD: \newline
Formalism and Implementation\label{sec:implement}}

The quark-parton picture is based on the factorization theorem of PQCD. The
conventional proof of the factorization theorem proceeds from the zero-mass
limit for all the partons---a good approximation at energy scales
(generically designated by $Q$) far above all quark mass thresholds
(designated by $M_{i}$). This clearly does not hold when $Q/M_{i}$ is of
order 1.\footnote{%
Heavy quarks, by definition, have $M_{i}\gg \Lambda _{QCD}$. \ Hence we
always assume $Q,M_{i}\gg \Lambda _{QCD}$.} It has been recognized since the
mid-1980's that a consistent treatment of heavy quarks in PQCD over the full
energy range from $Q\lesssim M_{i}$ to $Q\gg M_{i}$ can be formulated \cite%
{ColTun}. (This is most clearly seen in the CWZ renormalization scheme \cite%
{CWZ}.) The basic physics ideas were further developed in \cite{Acot2}; this
approach has become generally known as the ACOT scheme. In 1998, Collins
gave a general proof (order-by-order to all orders of perturbation theory)
of the factorization theorem that is valid for non-zero quark masses \cite%
{Collins98}. The resulting theoretical framework is conceptually simple: it
represents a straightforward generalization of the conventional zero-mass
(ZM) modified minimal subtraction (\msbar) formalism. This general mass (GM)
formalism is what we shall adopt.

The implementation of the general mass formalism requires attention to a
number of details, both kinematical and dynamical, that can affect the
accuracy of the calculation. Physical considerations are important to ensure
that the right choices are made between perturbatively equivalent
alternatives that may produce noticeable differences in practical
applications. We now systematically describe these considerations, and spell
out the specifics of the new implementation used in our study. \ For
simplicity, we shall often focus on the charm quark, and consider the
relevant issues relating to the calculation of structure functions at a
renormalization and factorization scale $\mu $ (usually chosen to be equal to
$Q$) in the neighborhood of the charm mass $M_{c}$. The same considerations
apply to the other heavy quarks, and to the calculation of cross sections.

\subsection{The Factorization Formula}

Let the total inclusive differential cross section for a general DIS
scattering process be written as
\begin{equation}
\frac{d^{2}\sigma }{dxdy}=N\sum_{\lambda }\,L_{\lambda }(x,y)F^{\lambda
}(x,Q)  \label{xSec1a}
\end{equation}%
where $F^{\lambda }(x,Q)$ are structure functions representing the forward
Compton amplitudes for the exchanged vector bosons on the target nucleon, $%
L_{\lambda }(x,y)$ are kinematic factors originating from the (calculable)
lepton scattering vertices, and $N$ is an overall factor dependent on the
particular process.\footnote{%
The summation index $\lambda $ can represent either the conventional tensor
indices $\left\{ 1,2,3\right\} $ or the helicity labels $\left\{
Right,Left,Longitudinal\right\} $. In the zero-mass case, these are the only
independent structure functions. However, in the general mass case, there
are five independent hadronic structure functions. In addition to the usual
two parity (and chirality) conserving, say $F_{2}$ and $F_{long}$, and one
parity violating, $F_{3}$, structure functions, there are also two
chirality-violating amplitudes (one each for $F_{2}$ and $F_{long}$). These
are proportional to $g_{R}g_{L}$, where $g_{R,L}$ are the electroweak
couplings of the vector boson to the quark line to which it is attached.
\cite{Acot2}} We follow the notation of Ref.\thinspace \cite{Acot1}, from
which detailed formulas for the above factors can be found. The PQCD
factorization theorem for the structure functions has the general form
\begin{equation}
F^{\lambda }(x,Q^{2})=\sum_{a}f^{a}\otimes \widehat{\omega }_{a}^{\lambda
}=\sum_{a}\int_{\zeta }^{1}{\frac{d\xi }{\xi }}\ f^{a}(\xi ,\mu )\ \widehat{%
\omega }_{a}^{\lambda }\left( \frac{x}{\xi },\frac{Q}{\mu },\frac{M_{i}}{\mu
},\alpha _{s}(\mu )\right) .  \label{master}
\end{equation}%
Here, the summation is over the active parton flavor label $a$, $f^{a}(x,\mu
)$ are the parton distributions at the factorization scale $\mu $, $\widehat{%
\omega }_{a}^{\lambda }$ are the Wilson coefficients (or
hard-scattering amplitudes) that can be calculated order-by-order in
perturbation theory, and we have implicitly set the renormalization
and factorization scales to be the same. The lower limit of the
convolution integral $\zeta $ is usually taken to be
$x=Q^{2}/2q\cdot p$---the Bjorken $x$---in the conventional ZM
formalism; but this choice needs to be re-considered when the Wilson
coefficients include heavy quark mass effects, as we shall do in
Sec.\thinspace \ref{sec:rescaling} below. In most applications, it
is convenient to choose $\mu =Q$; but there are circumstances in
which a different choice becomes useful. For DIS at order $\alpha
_{s}$ and beyond, the results are known to be quite insensitive to
the choice of $\mu $.

\subsection{The (Scheme-dependent) Parton Distributions and Summation over Parton Flavors\label%
{sec:suma}}

The summation $\sum_{a}$ over \textquotedblleft parton
flavor\textquotedblright\ label $a$ in the factorization formula, Eq.\,(\ref%
{master}), is determined by the \emph{factorization scheme} chosen
to define the Parton Distributions $f_a(x,\mu)$.

In the \emph{fixed flavor number scheme} (FFNS), one sums over $a=g,u,\bar{u}%
,d,\bar{d},...$ up to $n_{f}$ flavors of quarks, where $n_{f}$ is held at a
fixed value ($3,4,...$). For a given $n_{f}$ (say $n_f=3)$, the $n_{f}$-FFNS
has only a limited range of applicability, since, at order $m$ of the
perturbative expansion, the Wilson coefficients contain logarithm terms of
the form $\alpha _{s}^{m}\ln ^{m}(Q/M_{i})$ for $i>n_{f}$, which is not
infrared safe as $Q$ becomes large compared to the renormalized mass of the
heavy quark $M_{i}$ (e.g.\,$M_{c,b}$, in the $n_f=3$ example).

The more general \emph{variable flavor number scheme} (VFNS), as defined in
Collins' general framework \cite{ColTun,Collins98}, is really a \emph{%
composite scheme}: it consists of a series of FFNS's matched at conveniently
chosen \emph{match points} $\mu _{i}$, one for each of the heavy quark
thresholds. At $\mu _{i}$, the $n_{f}$-flavor scheme is matched to the $%
(n_{f}+1)$-flavor scheme by a set of perturbatively calculable finite
renormalizations of the coupling parameter $\alpha _{s}$, the mass
parameters $\left\{ M_{i}\right\} $, and the parton distribution functions $%
\left\{ f^{a}\right\} $. The matching scale $\mu _{i}$ can, in principle, be
chosen to be any value, as long as it is of the order of $M_{i}$. In
practice, it is usually chosen to be $\mu _{i}=M_{i}$, since it has been
known that, in the commonly used (VFNS) \msbar{} scheme, all the renormalized
quantities mentioned above are continuous at this point up to NLO in the
perturbative expansion \cite{ColTun}. Normally, when the VFNS is applied at a
factorization scale $\mu $ that lies in the interval ($\mu
_{n_{f}},\mu_{n_{f}+1}$), the $n_{f}$-flavor scheme is used.
Thus the \emph{number of active parton flavors} depends on the scale $\mu $%
---henceforth denoted by $n_{f}(\mu )$---and it increases by one when $\mu $
crosses the threshold $\mu _{i+1}$ from below.\footnote{%
Since, once properly matched, all the component FFNS's are well defined, and
can co-exist at any scale, it is possible to choose the transition from the $%
n_{f}$- to the ($n_{f}+1$)-flavor scheme at a \emph{transition scale} that
is different from the \emph{matching scale }$\mu _{i}$. It is, arguably,
even desirable to make the transition at a scale that is several times the
natural matching scale $\mu _{i}=M_{i}$, since, physically, a heavy quark
behaves like a parton only at scales reasonably large compared to its mass
parameter. But, with the rescaling prescription that we shall adopt in Sec.\,%
\ref{sec:rescaling}, the same purpose can be achieved with less technical
complications.}

PQCD in the VFNS is free of large logarithms of the kind mentioned above for
the FFNS---it is infrared safe, and hence remains reliable, at all scales $%
\mu $ ($\sim Q$) $\gg \Lambda _{QCD}$. In this scheme, the range of
summation over \textquotedblleft $a$\textquotedblright\ in the factorization
formula, Eq.\,(\ref{master}), is $0,1,..., n_{f}(\mu)$, where $0$ represents
the gluon, $1$ represents $u$\,and\,$\bar{u}$, etc.

Our implementation of the general mass formalism includes both FFNS
and VFNS. In practice, however, for reasons already mentioned, we
shall mostly use the VFNS. The definition of parton distributions in
the scheme described above is exactly the same as that of previous
CTEQ PDFs, all being based on \cite{ColTun}. (It is also shared,
essentially, by most global analysis groups, e.g.~MRST
\cite{Thorne06VFNS}.) The improvements of the formalism over
previous analyses reside in the consistent and systematic treatment
of mass effects in the Wilson coefficients and in the phase space
integral of Eq.\,(\ref{master}) that we shall now describe and
clarify.

\subsection{The Summation over (Physical) Final-state Flavors\label{sec:sumb}%
}

For total inclusive structure functions, the factorization formula, Eq.\,(%
\ref{master}), contains an implicit summation over all possible quark
flavors in the final state. One can write,%
\begin{equation}
\hat{\omega}_{a}=\sum_{b}\hat{\omega}_{a}^{b}
\end{equation}%
where \textquotedblleft $b$\textquotedblright\ denotes final state
flavors, and $\hat{\omega}_{a}^{b}$ is the perturbatively calculable
hard cross section for an incoming parton \textquotedblleft
$a$\textquotedblright\ to produce a final state containing flavor
\textquotedblleft $b$ \textquotedblright. (Cf.\,the Feynman diagrams
contributing to the calculation of the hard cross section in
Sec.\,\ref{sec:sacot} below. A caveat on the definition of
$\hat{\omega}_{a}^{b}$, is described in Sec.\,\ref{sec:facscale}.)

It is important to emphasize that ``$b$'' labels quark flavors that can be
produced \emph{physically} in the final state; it is \emph{not} a parton
label in the sense of initial-state parton flavors described in the previous
subsection. The latter (labelled $a$) is a theoretical construct and
scheme-dependent (e.g.\,it is fixed at three for the 3-flavor scheme);
whereas the final-state sum (over $b$) is over \emph{all flavors} that can
be physically produced. The initial state parton \textquotedblleft $a$%
\textquotedblright\ does not have to be on the mass-shell. But the final
state particles \textquotedblleft $b$\textquotedblright\ should be
on-mass-shell in order to satisfy the correct kinematic constraints and
yield physically meaningful results.\footnote{%
Strict kinematics would require putting the produced heavy flavor mesons or
baryons on the mass shell. In the PQCD formalism, we adopt the approximation
of using on-shell final state heavy quarks in the underlying partonic
process.} Thus, in implementing the summation over final states, the most
relevant scale is $W$---the CM energy of the virtual Compton process---in
contrast to the scale $Q$ that controls the initial state summation over
parton flavors (see next subsection).

The distinction between the two summations is absent in the simplest
implementation of the conventional (i.e., textbook) zero-mass parton
formalism: if all quark masses are set to zero to begin with, then all
flavors can be produced in the final state. \ This distinction becomes
blurred in a zero-mass (ZM) VFNS---the one commonly used in the literature
(including previous CTEQ analyses)---where the number of effective parton
flavors is incremented as the scale parameter $\mu$ crosses a heavy quark
threshold, but other kinematic and dynamic mass effects are omitted. \ Thus,
the implementation of the ZM VFNS by different groups can be different,
depending on how the final-state summation is carried out. This detail is
usually not spelled out in the relevant papers.

It should be obvious that, in a proper implementation of the general mass
(GM) formalism, the distinction between the initial-state and final-state
summation must be unambiguously, and correctly, observed. \ For instance,
even in the 3-flavor regime (when $c$ and $b$ quarks are \emph{not counted as
partons}), the charm and bottom flavors still need to be counted in the final
state---at LO via $W^{+}+d/s\rightarrow c$ or $W^{-}+u/c\rightarrow b$,
and at NLO via the gluon-fusion processes such as $W^{+}+g\rightarrow \bar{s}%
+c$ or $\gamma +g\rightarrow c\bar{c}\,(b\bar{b})$, provided there is enough
CM energy to produce these particles.

This issue immediately suggests that one must also give careful
consideration to the proper treatment of the integral over the final-state
phase space and other kinematical effects in the problem.

\subsection{Kinematic Constraints and Rescaling\label{sec:rescaling}}

Once mass effects are taken into account, kinematic constraints have a
significant impact on the numerical results of the calculation; in fact,
they represent the dominant factor in the threshold regions of the phase
space. \ In DIS, with heavy flavor produced in the final state, the simplest
kinematic constraint that comes to mind is
\begin{equation}
W-M_{N}>\sum _{f}~M_{f}  \label{KinConstraint}
\end{equation}%
where $W$ is the CM energy of the vector-boson--nucleon scattering process, $%
M_{N}$ is the nucleon mass, and the right-hand side is the sum of \emph{all}
masses in the final state. Since $W$ is related to the familiar kinematic
variables ($x,Q$) by $W^{2}-M_{N}^{2}=Q^{2}(1-x)/x$, this constraint can be
imposed by a step function $\theta (W-M_{N}-\sum_{f}M_{f})$ condition on the
right-hand side of Eq.\,(\ref{master}), irrespective of how, or whether, mass
effects are incorporated in the convolution integral. Although that simple
approach would represent an improvement over ignoring the kinematic
constraint Eq.\,(\ref{KinConstraint}), it is too crude, and can lead to
undesirable discontinuities.

A much better physically motivated approach is based on the idea of
rescaling. The simplest example is given by charm production in the LO CC
process $W+s\rightarrow c$. It is well-known that, when the final state charm
quark is put on the mass shell, the appropriate momentum fraction variable
for the incoming strange parton, $\chi $ in Eq.\,(\ref{master}), is not the
Bjorken $x$, but rather $\chi =x(1+M_{c}^{2}/Q^{2})$ \cite{Barnett76}. This
is commonly called the \emph{rescaling variable}.

The generalization of this idea to the more prevalent case of NC processes,
say $\gamma /Z+c\rightarrow c$ (or any other heavy quark), took a long time
to emerge \cite{AcotChi}, because this partonic process implies the
existence of a \emph{hidden heavy particle}---the $\bar{c}$---in the target
fragment. The key observation was, heavy objects buried in the target
fragment are still a part of the final state, hence must be included in the
phase space constraint, Eq.\,(\ref{KinConstraint}). Taking this effect into
account, and expanding to the more general case of $\gamma /Z+c\rightarrow
c+X$, where $X$ contains only light particles, it was proposed that the
convolution integral in Eq.\,(\ref{master}) should be over the momentum
fraction range $\chi _{c}<\xi <1$, where%
\begin{equation}
\chi _{c}=x\left( 1+\frac{4M_{c}^{2}}{Q^{2}}\right) \ \ .  \label{rescaling}
\end{equation}%
In the most general case where there are any number of heavy particles in
the final state, the corresponding variable is (cf.\,Eq.\,(\ref%
{KinConstraint}))
\begin{equation}
\chi =x\left( 1+\frac{\left( \Sigma _{f}~M_{f}\right) ^{2}}{Q^{2}}\right) \
\ .  \label{Rescaling}
\end{equation}%
This rescaling prescription has been referred to as ACOT$\chi$ in the recent
literature \cite{AcotChi,Thorne06VFNS}.

Fig.\,\ref{fig:Rescaling} helps to visualize the physical effects of
rescaling for charm production in NC DIS. \figRescaling In this plot, we show
constant $x$ and constant $\chi_{c}$ lines. The threshold for producing charm
corresponds to the line $W=2M_{c}$ (lower right corner), which coincides with
$\chi_{c}=1$. Far above threshold, $\chi_{c}\simeq x$. Close to the
threshold, $\chi_{c}$ can be substantially larger than $x$. For fixed (and
sufficiently large) $x$, as $Q$ increases (along a vertical line upward in
the plot, such as $x=0.1$), the threshold is crossed at
$Q^{2}=4M_{c}^{2}\,x/(1-x)$ (point C on the plot), beyond which $\chi _{c}$
decreases, approaching $x$ asymptotically (point D on the plot). For fixed
$Q$, as $x$ decreases from $1$ (along a horizontal line to the left), the
threshold is crossed at $x=\left( 1+{4M_{c}^{2}}/{\,Q^{2}}\right) ^{-1}$,
below which $\chi_{c}$ is shifted relative to $x$ according to
Eq.\,(\ref{rescaling}) or (\ref{Rescaling}).

Rescaling shifts the momentum variable in the parton distribution function
$f^{a}(\xi ,\mu )$ in Eq.\,(\ref{master}) to a higher value than in the
zero-mass case. For instance, at LO, the structure functions at a given point
A are proportional to $f(x,Q)$ in the ZM formalism; but, with ACOT$\chi $
rescaling, this becomes $f(\chi _{c},Q)$. The shift $x\rightarrow \chi _{c}$
is equivalent to moving point A to point B in Fig.\,\ref{fig:Rescaling}.

In the region where $\left( \Sigma _{f}\,M_{f}\right) ^{2}/Q^{2}$ is not too
small, especially when $f(\xi ,\mu )$ is a steep function of $\xi $, this
rescaling can substantially change the numerical result of the calculation.
\ It is straightforward to show that, when one approaches a given threshold (%
$M_{N}+\Sigma _{f}~M_{f}$) from above, the corresponding rescaling variable $%
\chi \rightarrow 1$. Since generally $f^{a}(\xi ,\mu )\longrightarrow 0$ as $%
\xi \rightarrow 1$, rescaling ensures a smoothly vanishing threshold
behavior for the contribution of the heavy quark production term to all
structure functions. This results in a universal\footnote{%
Since it is imposed on the (universal) parton distribution function part of
the factorization formula.}, and intuitively physical, realization of the
threshold kinematic constraint for all heavy flavor production processes.

\subsection{Hard Scattering Amplitudes and the SACOT Scheme\label{sec:sacot}}

The last quantity in the general formula Eq.\,(\ref{master}) that we need to
discuss is the hard scattering amplitude $\widehat{\omega }_{a}^{\lambda
}\left( x,\frac{Q}{\mu },\frac{M_{i}}{\mu },\alpha _{s}(\mu )\right) $.
These amplitudes are perturbatively calculable. To facilitate the
discussion, consider the special case of charm production in a neutral
current process. At LO and NLO, the Feynman diagrams that contain at least
one heavy quark ($c$ or $\bar{c}$) in the final state are depicted in Fig.\,%
\ref{fig:NloDiagm} for both the 3-flavor (lower) and 4-flavor (upper)
schemes. \figNloDiagm

For $\mu <M_{c}$, the 3-flavor scheme applies. In this scheme, there is no
charm parton in the initial state. The only diagram contributing to charm
production at order $\alpha _{s}$ is the gluon fusion diagram. If $M_{c}$ is
kept as nonzero, the hard scattering amplitude is finite; and the
calculation is relatively straightforward. The hard scattering amplitude
depends only on ($x,Q/M_{c}$), not on the factorization scale $\mu $.

For $\mu >M_{c}$, the 4-flavor scheme applies. The LO subprocess $\gamma
/Z+c\rightarrow c$, with a charm parton in the initial state, is of order $%
\alpha _{s}^{0}$. It represents the resummed result of collinear
singularities of the form $\alpha _{s}^{n}\ln ^{n}(\mu /M_{c})$, $n=1,2,...$%
, from Feynman diagrams of all orders in $n$. Since the $\alpha _{s}\ln (\mu
/M_{c})$ terms due to the NLO diagrams shown in Fig.\,\ref{fig:NloDiagm} are
already included in the resummation, they must be subtracted to avoid
double-counting. These are denoted by the \textquotedblleft subtraction
terms\textquotedblright\ in Fig.\,\ref{fig:NloDiagm}. The subtraction term
associated with the gluon-initiated diagram is of the form $\alpha _{s}\ln
(\mu /M_{c})\omega ^{0}(M_{c})\int_{\chi }^{1}(d\xi/\xi) \,g(\xi ,\mu
)P_{gq}(\chi /\xi )$ where $\omega ^{0}(M_{c})$ is the LO hard scattering
amplitude and $P_{gq}$ is the $g\rightarrow q$ splitting function of QCD
evolution at order $\alpha _{s}$. The specific choices adopted for $\chi $, $%
\mu $, and the $M_{c}$ dependence of $\omega ^{0}(M_{c})$ together precisely
define the chosen factorization scheme. To be consistent, the same
prescription must be used in evaluating the LO term, which takes the form
$c(\chi ,\mu )\,\omega ^{0}(M_{c})$, where$c(\chi ,\mu )$ is the charm parton
distribution (cf.\,Eq.\,(\ref{charmscat}) in Appendix \ref{app:rescale}).

The relation between the choice of $\chi $ (in the form of the rescaling
variable $\chi _{c}$) and the proper treatment of kinematics was discussed in
the previous subsection (Cf.~also Appendix \ref{app:rescale}). The freedom
associated with the choice of the $M_{c}$ dependence of the hard scattering
amplitudes was discussed in Refs.\,\cite{Collins98,sAcot}. The simplest
choice that retains full accuracy can be stated succinctly as: keep the heavy
quark mass dependence in the Wilson coefficients for partonic subprocesses
with only light initial state partons ($g,u,d,s$); but use the zero-mass
Wilson coefficients for subprocesses that have an initial state heavy quark
($c,b$). This is known as the SACOT scheme. For the 4-flavor scheme to order
$\alpha _{s}$ (NLO), this calculational scheme entails: (a) keep the full
$M_{c}$ dependence of the gluon fusion subprocess; (b) for NC scattering
($\gamma /Z$ exchanges), set all quark masses to zero in the quark-initiated
subprocesses; and (c) for CC scattering ($W_{\pm }$ exchange), set the
initial-state quark masses to zero, but keep the final-state quark masses on
shell (Cf.\,\cite{Acot2,sAcot}).

\subsection{Choice of Factorization Scale\label{sec:facscale}}

The final choice that has to be made is that of factorization scale $\mu $,
which connects the (soft) parton distributions and the hard scattering
amplitude. Provided the matching between the LO and the subtraction terms
are correctly implemented as described above (Sec.\,\ref{sec:sacot}), the
difference due to different choices of $\mu $ is formally of one order
higher than that of the perturbative calculation---as long as $\mu $ is of
the same order of magnitude as the physical hard scattering scale, say $Q$.
\ However, in the threshold region, the scale dependence can be quite
sensitive to the treatment of kinematics because of heavy quark mass
effects. \ On the other hand, it was shown in Ref.\,\cite{AcotChi} that,
once the kinematics are handled correctly according to the ACOT$\chi $
prescription (Secs.\,\ref{sec:suma}--\ref{sec:rescaling}), the $\mu $
dependence of the overall calculation becomes very mild.

In the conventional ZM formalism, the natural choice of the hard scale (the
typical virtuality) for the DIS process is $Q$. Hence $\mu =Q$ is almost
universally used in all practical calculations. In the GM formalism, we
should re-examine the possible choices.

The total inclusive structure function $F_{i}^{tot}$ is infrared safe.
Consider the simple case of just one effective heavy flavor, charm
(i.e.\,below the bottom and top production thresholds),
\begin{equation}
F_{i}^{tot}=F_{i}^{light}+F_{i}^{c}\ ,  \label{totalinc}
\end{equation}%
for any given flavor-number scheme (i.e.\,3-flavor, 4-flavor, ... etc.). If
we use the same factorization scale $\mu $ for both terms, then the sum is
insensitive to the value of $\mu $---the logarithmic $\mu$-dependence of the
individual terms cancel each other. \ Since the right-hand side of Eq.\,(\ref%
{totalinc}) is dominated by the light-flavor term $F_{i}^{light}$, and the
natural choice of scale for this term is $\mu =Q$, it is reasonable to use
this choice for both terms. \ This turns out to be a good choice in practice
as well, since the resulting $F_{i}^{tot}$ is then continuous across the
boundary separating the 3-flavor region ($\mu <M_{c}$) from the 4-flavor
region ($\mu >M_{c}$)---the line $Q=M_{c}$ in Fig.\,\ref{fig:Rescaling}.

Experimentally, the semi-inclusive DIS structure functions for producing a
charm particle in the final state $F_{i}^{c}$ is often presented.
Unfortunately, theoretically, $F_{i}^{c}(x,Q,M_{c})$ is \emph{not infrared
safe} beyond NLO. One may nonetheless perform comparison of NLO theory with
experiment with the understanding that the results are intrinsically less
reliable, and they can be sensitive to the choice of parameters. The
analytic expressions for $F_{i}^{c}$ in PQCD suggest that the typical
virtuality for this process is $\sqrt{Q^{2}+M_{c}^{2}}$ instead of $Q$. For
the factorization scale in this case, the choice $\mu =\sqrt{Q^{2}+M_{c}^{2}}
$ appears to be natural. This choice has the added advantage that $\mu >M_{c}
$ for all physical values of $Q$; hence, in practice, with this choice, one
stays always in the 4-flavor regime, avoiding the need to make a transition
from 3- to 4-flavor calculations when $Q$ crosses the value $M_{c}$,
cf.\,Fig.\,\ref{fig:Rescaling}.\footnote{%
There is no physical significance to the transition of $Q$ across the value $%
M_{c}$. The physical threshold for producing charm is at $W=2M_{c}$. The ACOT%
$\chi $ prescription ensures continuity across this threshold.}

%% file: text/2.difference.tex


\section{Differences between ZM and GM calculations\label{sec:difference}}

The GM version of the $n$-flavor scheme calculation reduces to the
conventional ZM one when the hard scale $Q$ is much larger than the quark
mass $M_{n}$. Thus differences between the two schemes are only expected to
be noticeable in the $Q\sim M_{n}$ region. Similarly, the differences between
the GM and ZM versions of the VFNS should occur mostly around the charm,
bottom and top threshold regions of the $(x,Q)$ plane. \

In general, among the various mass effects described in the previous section,
the most significant one numerically is that due to rescaling, $f(x,\mu
)\rightarrow f(\chi ,\mu )$. The size of this effect depends on: (i) the size
of the shift $x\rightarrow \chi $; and (ii) the rate of change of $f(x,\mu )$
at the relevant value of $x$. \ As can be seen from
Fig.\,\ref{fig:Rescaling}, the size of the shift  $x\rightarrow \chi $ is
largest when $Q\sim M_{n}$. According to (ii) above, however, this effect
will be significant only when  $f(x,\mu )$ is large and rapidly varying in
$x$. As we shall see below, this effect shows up most prominantly at small
$x$.

In the left panel of Fig.\,\ref{fig:GmZmDiff}, we show the fractional
differences between the GM and ZM calculations for $F_{2}^{\gamma }(x,Q)$
over the $(x,Q)$ plane. The magnitude of the fractional difference (in
percentage) is represented by the color coding shown along the right vertical
axis, and by the dashed contour lines. \ The light solid curves are constant
$\chi $ lines, taking into account the $c$ and $b$ quark masses. The
kinematic boundary (blue line) corresponds to the HERA energy reach.
We see that the expectations of the previous paragraph are borne out.%
\figGmZmDiff

Mass effects can be more readily seen for physical quantities that vanish in
the ZM limit. The most obvious example is the longitudinal structure function
in DIS, $F_{L}(x,Q)$, which vanishes at LO in the ZM formalism. It is
therefore useful to examine the importance of mass effects in $F_{L}$
quantitatively. In the right panel of Fig.\,\ref{fig:GmZmDiff}, we show the
fractional differences between the GM and ZM calculations for $F_{L}^{\gamma
}(x,Q)$ over the ($x,Q$) plane. \ Compared to the $F_{2}$ case, we see that,
in addition to the effects of rescaling, the mass effects in the hard
scattering is quite prominent. (Notice the different vertical scales of the
two plots.) We see that the differences are more noticeable and spread wider
in the charm and bottom threshold region than for the $F_{2}$ case, due to
the additional mass effects in the hard scattering amplitude.

The differences between the GM and ZM calculations demonstrated here will
have an impact on the global QCD analysis of PDFs, since the precision DIS
data sets from both fixed-target and HERA cover the kinematic region
highlighted in the above plots. Going from a ZM to a GM global analysis, the
PDFs will undergo some re-alignment among themselves. And, for reasons
described above, noticeable differences in the predictions for $F_{L}$ are
expected.

%% file: text/3.globalfit.tex


\section{New Global Analysis\label{sec:globalfit}}

We now apply the improved implementation of the GM formalism to the global
QCD analysis of the full HERA I DIS cross section data sets (cf.~next
subsection), along with fixed-target DIS, Drell-Yan (DY) data sets and
Tevatron Run I inclusive jet production data sets that were used in previous
CTEQ global PDF studies, in order to obtain the most precisely determined
parton distributions possible.

\subsection{Input to the Analysis}

Previous CTEQ global analyses of PDFs used \emph{structure function} data
for all available DIS experiments. By now, both H1 and ZEUS experiments have
published detailed \emph{cross section} data from the HERA I runs (1994 -
2000) for both NC and CC processes. We are able to use these cross
section data directly in the global analysis, so that the new analysis is
free of the model-dependent assumptions that usually go into the extraction
of structure
functions. This is important since, in addition to the dominant $%
F_{2}^{\gamma ,\gamma Z}$, we can also gain model-independent information on
the longitudinal and parity violating structure functions $F_{L}^{\gamma
,\gamma Z}$ and $F_{3}^{\gamma Z}$ from this more comprehensive study. For
this new effort to yield more accurate PDFs, and to produce more reliable
predictions on the various structure functions mentioned above, it is
crucial to use the available correlated systematic errors in the global
analysis, as we shall discuss below.

The HERA I cross section data sets that are included in this
analysis consist of the total inclusive NC and CC DIS processes, as
well as the semi-inclusive DIS processes with tagged final state
charm and bottom mesons. They are listed in Table I. %
\tblHoneData%

These data are supplemented by fixed-target and hadron collider data
sets used in the previous CTEQ global fits: BCDMS, NMC, CCFR, E605
(DY), E866 proton-deuteron DY ratio, CDF $W$-lepton asymmetry, and
CDF/D0 inclusive jet production. Details and references to these can
be found in Ref.\,\cite{Cteq6}.
We adopt the same $Q$- and $W$- cuts on experimental data as in Ref.\,\cite%
{Cteq6}; and the stability of our results with respect to varying
these cuts has been studied and reported in
Ref.\,\cite{Msu05Stability}.

\subsection{Parametrization of non-perturbative initial PDFs\label{sec:param}}

The parametrization of the non-perturbative parton distribution functions is
an important aspect of global QCD analysis since the robustness and
reliability of the resulting PDFs depends on a delicate balance between
allowing enough flexibility in the functional forms adopted to represent the
unknown physics on the one hand, and avoiding over-parametrization that
exceeds the constraining power of the available experimental input on the
other.

For this round of analysis, we have carefully re-examined this issue, and
tried a variety of functional forms, including exploring the number of degrees
of freedom associated with each parton flavor that current experimental data
can constrain. (A more detailed discussion is given in
Sec.\,\ref{sec:uncertainty} on quantifying uncertainties). Results that are
common to many reasonable choices of parametrization are considered
trustworthy; the most representative among the stable fits are then chosen as
the new standards. This effort results in some minor streamlining of the
parametrization used in previous CTEQ analyses \cite{Cteq6}. Details are
given in Appendix \ref{app:param}.

As with the previous analysis, we choose the input scale of $\mu _{0}=1.3$
GeV (which is also the charm quark mass, $M_c$, used in our calculations).
We assume the strange and anti-strange quarks are equal to each other (%
$s=\bar{s}$), and are proportional to the non-strange sea combination $(\bar{%
u}+\bar{d})$ at $\mu_{0}$. We find that, within the current global
analysis setup, the proportionality constant $\kappa$ (defined as
$(s+\bar{s})/(\bar{u}+\bar{d})$) is only weakly constrained. For the
purpose of the current analysis, we use the common value of $\kappa
=0.5$ (at $\mu _{0}=1.3$ GeV) that is well within the allowed range.
Finally, as in all existing global analyses, we assume the $c$ and
$b$ distributions to be zero at the scale corresponding to their
masses, and are generated by QCD evolution above that.

\subsection{New Global Fits}

The new global fit with the improved theoretical calculation and more
extensive DIS data sets results in even better agreement between theory and
experiment than the previous fits of CTEQ6M/CTEQ6.1M \cite{Cteq6}, CTEQ6HQ
\cite{Cteq6Hq}, and CTEQ6AB \cite{Cteq6AB}.%
\footnote{%
In terms of the overall $\chi^2$, used as a measure of the
goodness-of-fit in our global analysis, the decrease is $\Delta\chi^2
\sim 200$ for $2676$ data points when the same new data sets are
fitted with the ZM theory vs.~the GM theory. This is a significant improvement.
} %
This is an important confirmation of the standard model (SM) in
general, and the general mass PQCD formalism described in
Sec.\,\ref{sec:implement} in particular.

We choose a new central fit, designated as CTEQ6.5M, and 40 sets of
eigenvector PDFs that form an orthonormal basis characterizing
estimated uncertainties in the parton parameter space according to
the Hessian method described in \cite{MsuHes,Cteq6}. We shall
describe the characteristics of the central fit in this section, and
the full eigenvector sets in a later section on uncertainties
(Sec.\,\ref{sec:uncertainty}).

The experimental data that have the most influence on the
determination of the PDFs are, as is well known, the precision DIS
total inclusive measurements, along with the DY experiments (sea
quarks), $W$ lepton asymmetry measurements (flavor differentiation),
and collider inclusive jet production measurements (gluon). New to
the current global analysis (in addition to the replacement of NC
structure functions by the more complete cross section data) are the
HERA CC total inclusive and NC tagged heavy flavor inclusive
structure functions and cross sections.%
\footnote{We remark, as mentioned in Sec.\,\ref{sec:facscale}, the
theoretical underpinning for the semi-inclusive heavy quark
production process is less firm than for the total inclusive ones.
But it does hold at order $\alpha_s$---the order at which we perform
this analysis. Incidentally, we found that the \emph{natural} choice
of scale for heavy flavor production cross section calculation
mentioned in Sec.\,\ref{sec:facscale} leads to a noticeable, but
marginally significant, improvement in the overall
global fit (compared to the default $\mu=Q$).} %
These new data sets are fit quite well in the new round of global
analyses, demonstrating the consistency of the underlying PQCD
framework. However, due to the limited statistics available for
these processes, they do not provide any readily identifiable
constraints within the confines of the general global analysis
procedure. More dedicated studies, with targeted techniques, may be
needed to uncover potential physical implications of these data and
their HERA II successors. We shall not pursue these in this paper.

As examples of the new fits to the HERA I cross section data, we show two
plots comparing the H1 and ZEUS 1999-2000 $e^{+}p$ NC reduced cross section
data sets with the CTEQ6.5M fit. \figHoneTh%
The $\chi ^{2}/$Npts (number of data points) for the two data sets
are 169/147 and 94/90 respectively. The H1 data set has 6 sources of
correlated systematic errors $\{r_{i},i=1,6\}$. The optimal shifts
of these errors for the CTEQ6.5M fit are $\{r_{i}\} = \{0.378$,
$-0.173$, $0.413$, $0.329$, $0.544$, $-0.515\}$---all within one
$\sigma$, and $\sum_{i=1}^{6}r_{i}^{2}=1.012$. (The definition of
the $r_i$ parameters, as used in our treatment of correlated errors,
can be found in Appendix B of \cite{Cteq6}.) The ZEUS data set has 8
correlated systematic errors. The corresponding shifts are
$\{r_{i}\} = \{0.096$, $-0.110$, $-0.048$, $0.385$, $-0.068$,
$-0.651$, $-0.376$, $-1.334\}$---with only the last one being above
1, and $\sum_{i=1}^{8}r_{i}^{2}=2.522$. Both are quite reasonable.
This pattern is typical for other HERA data sets. Details are
available upon request to the authors.

%% file: text/4.comparison.tex


\subsection{New Parton Distributions \label{sec:NewPdf}}

Since both the updates in theory and in experimental input in this global
analysis represent incremental improvements over the previous CTEQ effort,
rather than major modifications, we do not expect drastic shifts in the
resulting PDFs. In the following discussions, we will focus on the few
notable differences and their physical implications.

To highlight the changes in the PDFs, we present the ratio of the new
CTEQ6.5M distributions and the corresponding CTEQ6.1M ones, compared to the
previously estimated uncertainty bands of the latter. Figure\,\ref%
{fig:PdfNewOld} shows the d-quark, u-quark and gluon distributions at $Q=2$
GeV. The CTEQ6.5M/CTEQ6.1M ratios are represented by the solid curves.
\figPdfNewOld \ To illustrate the universal behavior of the new fits, we also
include two dashed curves in each plot, representing equivalent good fits
with alternative parametrizations mentioned earlier (second paragraph of
Sec.\,\ref{sec:param}). We do not show separate results on the sea and the
valence distributions, since the $u$- and $d$-quark distributions are
dominated by the former at small $x$, and the latter at large $x$---both
visible in the existing plots.

We see from Fig.\,\ref{fig:PdfNewOld} that the new PDFs are indeed
generally within the previously estimated uncertainty bands,
demonstrating consistency with previous analyses. There is, however,
a notable departure of the new quark distributions from the old ones
in the region $x\sim 10^{-3}$. At the peaks of the ratio curves, the
new distributions are up to 20\% larger than CTEQ6.1M, and a factor
of two outside the previously estimated error bands. This feature is
shared by all choices of alternative parametrizations (and by all 40
sets of eigenvector PDFs to be discussed in the next section).

It is not surprising to see a shift in the extracted quark PDFs in the small-%
$x$ and low-$Q$ region, since this is where the theoretical treatment of
quark mass effect matters (cf.\,Sec.\,\ref{sec:difference}, particularly the
left plot of Fig.\,\ref{fig:GmZmDiff}). To see that this shift is indeed
caused by mass effects in the new calculation, we compare $F_{2}(x,Q^2=4$ GeV%
$^2)$ calculated using CTEQ6.5M PDFs with the general-mass and the
zero-mass Wilson coefficients, both normalized to the CTEQ6.1M
result (which was obtained in the ZM formalism) in
Fig.\,\ref{fig:F2comQ}. \figFtwocomQ The difference between the two
CTEQ6.5M calculations is broadly in the small-$x$
region, and it is of similar order of magnitude to that seen in Fig.\,\ref%
{fig:PdfNewOld}. Since the GM calculation (solid curve) yields lower values
than the ZM calculation (dashed curve), the new CTEQ6.5M quark PDFs are
pushed up in the new global analysis (cf.\,Fig.\,\ref{fig:PdfNewOld}) in
order to fit the same DIS data. The deviation of the GM CTEQ6.5M prediction
is not far from the ZM CTEQ6.1M result (horizontal reference line, $1.00$) in Fig.\,\ref%
{fig:F2comQ} since both are obtained by fits to the data. However,
they are not identical because there are several differences between
the old and new global fits---DIS cross sections vs.~structure
functions as input, parametrization, etc.---in addition to the
treatment of heavy quark mass effects.

The heavy quark mass effects diminish with increasing $Q^2$. However, their
effect on the analysis of experiments at low $Q^2$ produces a change in the
PDFs even at larger $Q^2$. \figQfive Figure \ref{fig:Qfive} shows the same
comparisons at the scale $Q^2=25 \, \mathrm{GeV}^2$. We see that the
difference between CTEQ6.5M and CTEQ6.1M remains significant at this scale.
Even at $Q^2 \sim 10^4 \, \mathrm{GeV}^2$, the CTEQ6.5 quark PDFs are higher
than CTEQ6.1 by $\sim 5-6\%$ for $x \sim 10^{-3}$---approximately the CTEQ6.1
uncertainty at these $(x,Q)$ values. This can result in significant increases
in physical predictions on hadron collider cross sections that are sensitive
to PDFs in this $x$ range, e.g.~for $W/Z$ production at the LHC, the increase
is $\sim 8\%$, cf.\,Sec.\,\ref{sec:PhyPred}.

\subsection{Mass effects, Low-$Q^{2}$ HERA data, and Correlated systematic
errors}

The noticeable changes in the quark distributions at low $x$ and $Q$
also suggest a closer examination of the comparison of the new
theoretical predictions with the precision DIS data in this region,
particularly because the longitudinal structure function is expected
to play a substantive role in the understanding of the low $x$ and
$Q$ HERA cross section data.

Among the high precision HERA I data sets, the 1996-97 $e^{+}p$ NC
reduced cross section measurements include data in the low $x$ and
$Q$ region. We examine these data sets (from H1 \cite{Adloff:2000qk}
and ZEUS \cite{Chekanov:2001qu} ) in a little more detail. The
CTEQ6.5 fit to the H1 data set has
$\chi ^{2}/\mathrm{Npts} = 107\,/\,115$; the shifts of the 5 correlated
systematic errors are
$\{r_{i}\} = \{0.218$, $1.361$, $-0.472$, $0.374$, $1.581\}$
with $\sum r_{i}^{2}=4.763$. The corresponding numbers for the ZEUS
data set are $\chi^{2}/\mathrm{Npts}= 279\,/\,227$; the shifts of the
10 correlated systematic errors are
$\{r_{i}\} = \{-1.575$, $-0.573$, $-1.407$, $-0.263$, $-0.025$,
$-1.203$, $1.278$, $0.425$, $-0.258$, $0.238\}$,
with $\sum r_{i}^{2}=8.244$. Aside from the somewhat high overall $\chi^2$
for the ZEUS data set (which was also seen in the previous round of CTEQ6.1
analysis using the corresponding $F_2$ data set), these numbers indicate
reasonable fits.

In Fig.\,\ref{fig:LowQHera}, we show the data from each experiment from the
four lowest $Q$-bins that pass our $Q > 2 \, \mathrm{GeV}$ cut, normalized to
the CTEQ6.5M fit. \figLowQHera The comparison with H1 data (left plot) shows
a pattern of \textquotedblleft turnover" of experimental data points (open
circles) at low $x$ with respect to the theory for all four $Q^{2}$ bins.
(For given $E$ and $Q$, low $x$ corresponds to high $y$.) However, this
discrepancy disappears when correlated systematic errors are included in the
analysis, as seen from the fact that the solid dots (representing data
corrected by systematic errors) fit rather well the theory predictions (the
horizontal lines corresponding to $1.00$ for the ratio). The values for the
overall $\chi ^{2}$ of this fit as well as the systematic shifts given in the
previous paragraph support this observation.

The comparison of the ZEUS data to the CTEQ6.5M fit (right plot) does not
show the same systematic low-$x$ turnover. Instead, data (open circles) in
the three higher $Q$ bins are generally below the theory prediction. Again,
we see that the differences between the two go away when correlated
systematic errors are included in the analysis (solid dots). We get
acceptable fits in all bins with reasonable systematic shifts.

The low-$Q$ and high-$y$ HERA data has been the subject of special
analyses by both H1 and ZEUS collaborations, mainly in the context
of extracting the longitudinal structure function $F_{L}(x,Q)$.
Results of QCD fits performed in this regard
\cite{Adloff:2000qk,H102Fl}, appear to be similar to those shown
above. Detailed comparison, however, is not possible at present
since details of the theoretical input to the HERA analyses (e.g.\
issues related to mass effects described in
Sec.\,\ref{sec:implement}) has not been specified. As we have shown,
heavy quark mass effects are important in this kinematic region. \
It would be useful for all future analyses to include the mass
effects.

The observed low-$x$ turnover of the H1 data has been considered a potential
problem for global analyses by the MRST group \cite{Thorne05Fl,MST06Fl}. This
difficulty does not seem to arise in our analysis. Two possible sources could
be responsible for this difference: (i)  although both analyses
include quark mass effects, the implementations are not the same (cf.\,\cite%
{Thorne06VFNS}, compared to Sec.\,\ref{sec:implement}); (ii) our
inclusion of correlated systematic errors in the analysis is
responsible for bringing theory and experiment into agreement, as
demonstrated in Fig.\,\ref{fig:LowQHera}. A more detailed study of
these issues is clearly called for.

%% file: text/5.uncertainty.tex


\section{Uncertainties on New Parton Distributions\label{sec:uncertainty}}

Using the new theory implementation and experimental input, we have also
performed a detailed study of the uncertainties of the PDFs, following the
Hessian method of \cite{MsuMV,MsuHes,MsuLgr,Cteq6}. \ This involves finding
a set of eigenvector PDFs that characterize the uncertainties of the parton
distributions around the \textquotedblleft best fit\textquotedblright\ in
the parton parameter space.

To ensure that this procedure will yield meaningful results, we first carry
out a series of studies to match our fitting parametrizations with
theoretical and experimental constraints: (i) first we make sure that the
best fit is robust by checking that the quality of the global fit cannot be
significantly improved by increasing the number of free parameters or
changing the functional forms; (ii) next, we identify \textquotedblleft flat
directions\textquotedblright\ in the resulting parameter space (representing
degrees of freedom that are not constrained by current experimental input)
by diagonalizing the Hessian matrix and examining its variation along the
eigenvector directions; (iii) using that information, we freeze an
appropriate subset of parameters and re-diagonalize the Hessian matrix,
which characterizes the quadratic dependence of $\chi^{2}$ on the fitting
parameters in the neighborhood of the minimum. The number of eigenvectors
that can reasonably be determined is around 20---the same as was used in the
CTEQ6.1 analysis.

To arrive at a quantitative estimate of the range of uncertainties in the
parton parameter space, we examine the global $\chi _{\mathrm{global}}^{2}$
that is used by the fitting program as a measure of the overall
\textquotedblleft goodness-of-fit", as the parameters are varied in the
neighborhood of the global minimum. In defining $\chi _{\mathrm{global}}^{2}$%
, we include weight factors for a few experiments that have only a small
number of data points. The eigenvectors and the weights are arrived at by the
iterative method of Ref.\thinspace \cite{MsuMV} so that an adequate fit to
every data set is maintained as far out as possible along the eigenvector
directions. This allows us to estimate a confidence range by the condition
that, within this range, the fit to every data set is within its 90\%
confidence level.

In this way we generate the final eigenvector PDF sets so that they span the
\emph{90\% confidence range} for the contributing data sets. The procedure is
similar to that formulated in Refs.~\cite{MsuLgr,MsuHes,Cteq6} (which
contains more details). There are 40 eigenvector sets, corresponding to
displacements from the central fit in the ``$+$'' or ``$-$'' senses along
each of the eigenvector directions.\footnote{%
The uncertainty bands are not always symmetric in the +/- directions since
we independently generated +/- sets along each eigenvector direction in
order to provide somewhat better approximation to the uncertainties along
the flatter directions, where there are deviations from the quadratic
approximation.} PDF uncertainty limits for a \emph{90\% confidence range}
for physical predictions can be calculated from these sets by computing the
prediction for each of the 40 sets, and adding the approximately 20 upward
(downward) deviations in quadrature to obtain the upper (lower) limit.

\figPdfBandA 

Figure~\ref{fig:PdfBandA} shows the uncertainty bands determined by this
method for $u$, $d$, and $g$ PDFs at the scale $Q^2=4 \, \mathrm{GeV}^2$.
The lines represent each of the 40 sets of eigenvector PDFs normalized to
the central fit. The upper (lower) edges of the shaded uncertainty region is
obtained by adding in quadrature the contributions from eigenvector sets
that lie above (below) the central fit at each particular $x$. We observe
that, in some cases (such as for the gluon at large $x$), the uncertainty
band is dominated by a single pair of eigenvector PDFs, corresponding to the
two senses along a single eigenvector direction.

\figPdfBandB

Figure~\ref{fig:PdfBandB} shows the same uncertainty bands as above,
together with curves that show the fractional uncertainty range from CTEQ6.1
that was determined by a similar procedure. This shows that the two
estimates of PDF uncertainties are broadly comparable with each other, with
a slight tightening of the uncertainty ranges of $d$ quark and gluon
distributions in certain $x$ regions. Of more interest is a comparison of
estimated uncertainties of physical predictions. We shall discuss these in
Sec.\,\ref{sec:PhyPred} below.

Because of the improved theoretical and experimental input to this new
global analysis, as well as the much more thorough study of its
parametrization dependence, we now have greater confidence in these
uncertainty estimates than before. Wider applications of the new results to
Standard Model and New Physics processes at hadron colliders will be pursued.

To compare some currently used PDFs to that of CTEQ6.5, Figure~\ref%
{fig:PdfBandC}\figPdfBandC shows CTEQ6.1M (dashed curves), CTEQ6A118
(central fit of the CTEQ6 ``$\alpha_s$ series'' \cite{Cteq6AB}), and MRST04
\cite{Mrst04PhyG} PDFs as ratios to CTEQ6.5M. We note that the previous CTEQ
PDFs lie mainly within the new uncertainty bands---except at $x \sim 10^{-3}$
for the reasons discussed in the various subsections of Sec.\,\ref%
{sec:globalfit}. The MRST04 quark PDFs are closer to CTEQ6.5 in the $x\sim
10^{-3}$ region than CTEQ6.1M and CTEQ6AB, presumably because both CTEQ6.5
and MRST04 include mass effects (albeit using different prescriptions) while
the other two are in the ZM formalism. The MRST04 gluon PDFs are within the
CTEQ6.5 uncertainty band at large-$x$; but they are outside the band around $%
x=0.3$ and at small $x$.

As mentioned earlier, as a part of our investigation on the robustness of
our results, we have performed uncertainty analysis using different number
of variable parameters. (Other global analysis groups have always used fewer
variables.) We found the resulting ranges of uncertainty stable if this
number is greater than 16. To show how the results may be affected by too
restrictive choices of parameters, Figure~\ref{fig:PdfBandC} also include
two curves (red) that correspond to the edges of the uncertainty bands using
only 11 free parameters. We see that these bands are considerably narrower
than the stable results represented by the CTEQ6.5 bands. In other words,
over-restricting the degrees of freedom in the input parametrization at $%
\mu_0$ can significantly overestimate how well the PDFs are measured.

%% file: text/6.summary.tex

\section{Implications for Hadron Collider Physics\label{sec:PhyPred}}

The new PDFs have significant implications for hadron collider
phenomenology at the Tevatron and the LHC. We shall mention one
example here: the benchmark $W$ production total cross section
$\sigma_W$.

The higher quark distributions in CTEQ6.5 for $x\sim 10^{-3}$ lead to an
increase in the predicted values for $\sigma_W$ over those based on CTEQ6.1:
for Tevatron Run II, we get $\Delta\sigma_W / \sigma_W = 3.5 \%$; and for the
LHC, $\Delta\sigma_W / \sigma_W = 8 \%$.%
\footnote{%
The predicted values for the total $W$ production cross section
using CTEQ6.5M is 24.7 nb at the Tevatron II, and 202 nb at the LHC.} %
The significant increase of the predicted  $\sigma_W$ at the LHC
reflects the fact that it is directly dependent on PDFs in the
region $x\sim 10^{-3}$.

It is also interesting to compare the uncertainties of these
predictions as estimated by the Hessian method. For the Tevatron, we
find this uncertainty to be $\mathrm{+}3.1/\mathrm{-}3.0\%$ for
CTEQ6.5, compared to $\mathrm{+}3.8/\mathrm{-}4.4\%$ estimated using
CTEQ6.1. For the LHC, the uncertainty is
$\mathrm{+}4.9/\mathrm{-}4.1\%$ for CTEQ6.5, compared to
$\mathrm{+}5.2/\mathrm{-}5.9\%$ for CTEQ6.1. Thus there is a notable
reduction in the uncertainty ranges. This is perhaps related to a
better determination of the correlations between different parton
flavors, in addition to the obvious connection to the uncertainties
of the individual PDFs (which are rather comparable for CTEQ6.1 and
CTEQ6.5) shown in Figure~\ref{fig:PdfBandB}.

Detailed results on W/Z production, including differential distributions, and
implications of these new PDFs on other SM and New Physics processes for
hadron colliders will be explored in a separate study.

\section{Summary and Outlook\label{sec:summary}}

We have updated both the theoretical and experimental input to the
CTEQ global QCD analysis in this work. On the theory side, we have
used a newly implemented systematic approach to PQCD, including
heavy quark mass effects according to the general formalism of
Collins \cite{Collins98}. Experimentally, we have used all available
precision HERA I cross section data sets, along with
well-established fixed target and hadron collider experimental data.
Correlated systematic errors, whenever available, are fully
incorporated in the analysis. In performing this analysis, we have
extensively studied the effects due to changes of the functional
form for the initial parton distributions, and to changes in the
number of free parameters allowed in the fits, in order to arrive at
stable and physically meaningful results.

One noticeable common feature of the new parton distributions, compared to
the previous ones, is the change in the $u$ and $d$ quark distributions
around $x\sim 10^{-3}$ at low $Q$.
This results from the inclusion of mass effects
in the theory, in conjunction with the improved (model independent) HERA
cross section data, over the previous analysis (using $F_{2}$ data that
inevitably involve assumption about $F_{L}$ for their extraction).

Within the general-mass PQCD framework, we find broad consistency between
the extensive data sets incorporated in this analysis.\footnote{%
Given the fact that several individual experimental data sets show
much larger fluctuations than expected from normal statistics, and
that problems of statistical compatibility between different
experiments of the same type are not uncommon, it is well-known that
this complex system is too un-textbook-like to be amenable to strict
``$1 \,\sigma$ error'' analysis of the PDF parameters. As mentioned
in the main body of this paper, we generally apply 90\% confidence
level ``goodness-of-fit" criteria in our uncertainty studies.} %
This permits us to arrive at 90\% confidence level
estimates of the uncertainties of PDFs around the chosen central fit,
CTEQ6.5M. These uncertainties are encapsulated in 40 sets of eigenvector PDFs
that span the neighborhood of the central fit in the parton parameter space.
We found the new uncertainty bands of the PDFs are slightly narrower than the
previous ones, but are generally of the same order of magnitude.

While progress towards better-determined PDFs in a more precisely
formulated PQCD framework has clearly been made, it is worthwhile to
mention some of the limitations of current global analysis of PDFs
that call for continued advances in both theory and experiment.
First, due to rather weak existing experimental constraints of the
strange quarks, we have assumed in this work that $(s+\bar{s})$ is
of the same shape as $(\bar{u}+\bar{d})$ and fixed the
proportionality constant $\kappa $ at the initial scale $Q_{0}$
during the fit. In principle, better constraints on $\kappa $, and
on the possible difference between $s$ and $\bar{s}$, can come from
recent neutrino scattering experiments by NuTeV \cite{NuTeVxsc} and
Chorus \cite{ChorusSFxsc}. However, many open questions pertaining
to the consistency between existing experimental data sets
(including those from the \textquotedblleft old" experiments CDHSW
and CCFR), nuclear target corrections, and other issues make the use
of these data controversial at present. A dedicated study on the
strangeness sector, designed to delineate the range of uncertainties
of both $s(x)$ and $\bar{s}(x)$, will require a somewhat different
tactic, focused on the most relevant degrees of freedom. In the same
vein, we have assumed that all heavy quark partons ($c$ and $b$) are
generated \emph{radiatively} by QCD evolution (mainly gluon
splitting) at the initial scale $Q_{0}$. This assumption is not well
defined quantitatively, because it depends on the choice of $Q_{0}$.
Furthermore the assumption itself may be questioned---does intrinsic
charm exist in the proton? The charm and bottom degrees of freedom
can be investigated phenomenologically within the general-mass PQCD
framework described here, and is currently under investigation
\cite{Wkt06Dis2}.

There has been considerable discussion in recent literature \cite{DIS06}
about extending global QCD analysis to higher orders in $\alpha_{s}$. This
interest is spurred, on the one hand, by the final availability of the NNLO
evolution kernel \cite{MVV04}, and on the other hand, by the calculation of
NNLO Wilson coefficients for various hard scattering processes, such as Higgs
production \cite{Anastasiou}. In the global analysis of PDFs, NNLO becomes
important, and, by implication, necessary to include, when the theoretical
corrections to NLO calculations are comparable to the errors on the
corresponding input experimental data. For DIS and DY processes used in
current global analysis, this happens only in the small-$x$ region
\cite{MVV04}. However, near boundaries of the kinematic region, such as small
$x$, the relatively large corrections are generally associated with higher
powers of large logarithms in PQCD. These are symptoms of the breakdown of
the fixed-order perturbation expansion and the need to resum these logarithms
in order to achieve stable and reliable results. Thus, efforts to expand
global QCD analysis to higher orders must go hand-in-hand with work to
incorporate resummation effects (not only confined to small $x$) in the
theoretical framework.%
\footnote{%
Much recent progress on small-$x$ resummation, and its possible
application the global QCD analysis, has been reported at the
DIS2006 Workshop, and reviewed in \cite{Thorne06DisThSum}. Other
types of resummation, such as transverse momentum and threshold
resummations, have also come to the fore because of their importance
for LHC phenomenology.
} %
The importance of much development work along both lines is evident.

The new CTEQ6.5 PDF sets, including the eigenvector sets discussed
in Sec.\,\ref{sec:uncertainty}, will be made available at the CTEQ
web site (http://cteq.org/) and through the LHAPDF system
(http://hepforge.cedar.ac.uk/lhapdf/). Because the GM formalism used
in this analysis represents a better approximation to QCD, compared
to the previously used ones, the new PDFs are expected to be closer
to the true values than previous ones. In physical applications at
energy scales much larger than $M_c$ and $M_b$, these PDFs can be
convoluted with commonly available hard-scattering cross sections
calculated in the ZM formalism to obtain reliable predictions,
because quark mass effects in the Wilson coefficients will be
negligible. However, for quantitative comparison of a theoretical
calculation to precision DIS data, in the region where $M_c/Q$ and
$M_b/Q$ are not very small, the GM formalism described in
Sec.\,\ref{sec:implement} must be applied to the hard scattering
cross section together with CTEQ6.5 PDFs in order to obtain accurate
results.

\paragraph{Acknowledgements}

We thank John Collins, Stefan Kretzer, and Carl Schmidt for discussions about
the general PQCD framework, particularly the implementation of the general
mass scheme; and Joey Huston for discussions on various issues about global
analysis, as well as on the presentation of our results.

This work was supported in part by the U.S. National Science Foundation under
awards PHY-0354838 and PHY-0555545; and by National Science Council of Taiwan
under grant 94(95)-2112-M-133-001.

%% file: text/a1.rescale.tex


\section{Rescaling}

\label{app:rescale}

We give here an intuitive derivation of the rescaling prescription (ACOT$\chi
$) in the context of NC charm production. The general idea is applicable to
other heavy flavor production processes, as will become clear later. As mass
effects are only relevant at energy scales comparable to the heavy quark mass
(charm in our case), we will focus in this region where the physically
dominant mechanism for charm production is the \emph{gluon fusion} process
$\gamma ^{\ast }g\rightarrow c\bar{c}$. The graphical
representation of this process is shown as the first term in Fig.\,\ref%
{fig:rescale}.\figRescale The analytic expression for this contribution to
the physical structure function is
\begin{equation}
{\alpha _{s}(\mu )\int_{\chi _{c}}^{1}\frac{dz}{z}g(z,\mu )\omega _{\gamma
^{\ast }g\rightarrow c\bar{c}}^{1}(\frac{\chi _{c}}{z},Q,M_{c})\rule%
{0em}{3ex}}  \label{gluonfusion}
\end{equation}%
where ${\alpha _{s}\omega _{\gamma ^{\ast }g\rightarrow
c\bar{c}}^{1}}$ is the order ${\alpha _{s}}$ partonic cross section,
${g(z,\mu )}$ is the gluon distribution function. The lower limit of
the convolution integral is determined
by the boundary of the final state phase space integration; it is%
\begin{equation}
\chi _{c}=x(1+4M_{c}^{2}/Q^{2})  \label{rescalinga}
\end{equation}%
with $x$ being the Bjorken $x$. We recognize that this is just the
rescaling variable Eq.\,(\ref{Rescaling}) of Sec.\,\ref{sec:rescaling}.
For the gluon-fusion subprocess, it follows
strictly from kinematics, once we include the mass of the heavy
quarks into consideration.

The gluon fusion term by itself is not infrared safe at high energies. \ In
this limit, the infrared unsafe part comes from the collinear configuration
in the final state phase space integration. The singular part of ${\omega
_{\gamma ^{\ast }g\rightarrow c\bar{c}}^{1}}$ is of the form%
\begin{equation}
{\omega _{\gamma ^{\ast }g\rightarrow c\bar{c}}^{1}(z,,Q,M_{c})\rightarrow
\ln (\frac{Q}{M_{c}})P_{g\rightarrow c}(z)\;\omega _{\gamma ^{\ast
}c\rightarrow c}^{0}} \label{ColinearSing}
\end{equation}%
where${\;\omega _{\gamma ^{\ast }c\rightarrow c}^{0}}$ represents the order $%
\alpha _{s}^{0}$ process $\gamma ^{\ast }c\rightarrow c$ (upper vertex of
vector-boson coupling to quark); and ${P_{g\rightarrow c}(z)}$ is the
$g\rightarrow c $ splitting function. This collinear singularity of the
$\gamma
^{\ast }g\rightarrow c\bar{c}$ contribution can be removed by a \emph{%
subtraction term} of the form%
\begin{equation}
-{\alpha _{s}(\mu )\ln (\frac{\mu }{M_{c}})\int_{\zeta }^{1}\frac{dz}{z}%
g(z,\mu )P_{g\rightarrow c}(\frac{\zeta }{z})\;\omega _{\gamma ^{\ast
}c\rightarrow c}^{0}}  \label{subtraction}
\end{equation}%
with the introduction of the factorization scale $\mu $ (generally chosen to
be of order $Q$). This term is represented by the second graph in Fig.\,\ref%
{fig:rescale}, where the mark on the internal charm parton line signifies
that its momentum $k$ is collinear to the gluon and $k^{2}\approx 0 $. For
the purpose of cancelling the collinear singularity of the gluon-fusion term
at high energies (the Bjorken limit), the variable $\zeta $ in this
expression can be any expression provided $\zeta \rightarrow x$ in that
limit. In fact, conventionally, it is taken to be just $x$. The trouble with
this choice is that in the other limit---near the threshold region where $W$
and $Q$ are of the order of $M_{c}$---the subtraction term knows nothing
about the kinematics of $c\bar{c}$ pair production, hence bears no relation
to the physical structure function. The combined result of
Eqs.~(\ref{gluonfusion}) and (\ref{subtraction}) is then completely
artificial, hence unphysical, in this region.  To remedy this problem, one
only needs to realize the
origin of the subtraction term, and make the obvious choice%
\begin{equation}
\zeta =\chi _{c}=x(1+4M_{c}^{2}/Q^{2})  \label{rescalingb}
\end{equation}%
that is appropriate for the parent gluon-fusion contribution,
Eq.\,(\ref{gluonfusion}). With this choice of the scaling variable
$\zeta$, the subtraction term, Eq.\,(\ref{subtraction}) behaves
correctly both in the high-energy and the low-energy regions.

To complete the derivation, we need to turn to the third diagram of Fig.\,%
\ref{fig:rescale}, which represents the simple order-$\alpha _{s}^{0}$
$\gamma ^{\ast }c\rightarrow c$ parton process in the 4-flavor scheme. From
the perspective of the preceding discussion, this term arises from resumming
the collinear and soft singularities to all orders in the perturbation expansion.
The leading term in this expansion is given by Eq.\,(\ref{subtraction}) above
(with a positive sign). The resummed result, as illustrated by the diagram,
is just
\begin{equation}
{c(\zeta ,\mu )\;\omega _{\gamma ^{\ast }c\rightarrow c}^{0}}
\label{charmscat}
\end{equation}%
Here, the choice of the scaling variable $\zeta $ should be dictated
by similar considerations as above: at high energies, $\zeta $ must
reduce to the Bjorken $x$; and in the threshold region, the combined
contribution from Eq.\,(\ref{charmscat}) and
the subtraction term, Eq.\,(\ref{subtraction}), must be of higher order in $%
\alpha _{s}$. The naive (and common) choice $\zeta =$ $x$ satisfies the first
criterion, but not the second. For the same reason discussed before, the
choice $\zeta =\chi _{c}$, Eq.\,(\ref{rescalingb}), satisfies both; hence it
is the  physically sensible one to use.

With the use of the rescaling variable $\zeta =\chi _{c}$ for the LO
$\gamma ^{\ast }c\rightarrow c$ term and the subtraction term, the
sum of the three contributions in Fig.\,\ref{fig:rescale} reduces,
by definition, to the gluon fusion contribution in the threshold
region, as it should; and it approaches the conventional zero-mass
PQCD form, LO ($\gamma ^{\ast }c\rightarrow c$) + (NLO correction),
in the high energy limit, as it should. From this perspective, one
sees clearly the dual role of the subtraction term: in the threshold
region, it overlaps substantially with the LO ($\gamma ^{\ast
}c\rightarrow c$) contribution to make the gluon fusion subprocess
the primary production mechanism; and in the high energy limit, it
overlaps with the singular part of the $\gamma ^{\ast }g\rightarrow
c\bar{c}$ contribution, and thus helps to render the combined order
$\alpha _{s}$ terms infrared safe (and yield the true NLO correction
to the perturbative expansion).

%% file: text/a2.param.tex


\section{Parametrization}

\label{app:param}

The parametrization of the parton distributions at $\mu_{0}$ that was used to
obtain the CTEQ5 and CTEQ6 parton distributions contained 5 shape parameters
(apart from normalization) for each flavor. However, the global analysis data
sets were not sufficiently constraining to determine all of these parameters,
so a number of them were frozen at some particular values.

In the current effort to match the theoretical parametrization with
experimental constraints, we achieve the same goal by adopting a
simpler form with 4 shape parameters for the valence quarks
$u_{v}(x)$, $d_{v}(x)$, and the gluon $g(x)$:
\begin{equation}
f(x)=a_{0}\,x^{a_{1}}\,(1-x)^{a_{2}}\,\mathrm{e}^{a_{3}x\,+\,a_{4}x^{2}}\;.
\label{eq:app1}
\end{equation}
This can be seen as a plausible generalization of the conventional
minimal form
\begin{equation}
f_{0}(x)=a_{0}\,x^{a_{1}}\,(1-x)^{a_{2}}\;,  \label{eq:app2}
\end{equation}
which combines Regge behavior at $x\rightarrow 0$ and
spectator counting behavior at $x\rightarrow 1$ in an economical way.

Both functions, (\ref{eq:app1}) and (\ref{eq:app2}), are conveniently
positive definite.
The following modified logarithmic derivatives of these functions are simple
polynomials in $x$,
\begin{equation}
\phi _{0}(x)=-x\,(1-x)\,\frac{d\ln f_{0}}{dx} = -a_{1}+b_{1}x\;,
\label{eq:app3}
\end{equation}%
and%
\begin{equation}
\phi (x)=-x\,(1-x)\,\frac{d\ln f}{dx}=-a_{1}+b_{1} x+
b_{2}x^{2}+b_{3}x^{3}\;,  \label{eq:app4}
\end{equation}%
where the coefficients $\left\{b_{i}\right\}$ are simple linear combinations
of the original ones $\left\{a_{i}\right\}$ in the exponent of
Eq.~(\ref{eq:app1}). So the form Eq.~(\ref{eq:app1}) corresponds to
generalizing the logarithmic derivative from a linear function $\phi_{0}(x)$
to a cubic polynomial $\phi(x)$, and follows in the spirit of using
polynomials to approximate unknown functions that have no known
singularities. The only practical question is whether this polynomial
generalization contains enough flexibility to represent the physical PDFs
that we are trying to determine. Our investigation indicates that this is the
case, since significantly better fits cannot be achieved by introducing additional
parameters or by changing the functional forms.

We continue to use the same parametrizations for $\bar{u}$, $\bar{d}$ that
were used in CTEQ6. As mentioned in the text (Sec.~\ref{sec:param}), we
continue to use the approximation $s(x)=\bar{s}(x)\propto \bar{u}(x)+\bar{d}%
(x)$. The full set of formulas for the initial PDFs at
$\mu_{0} = 1.3 \, \mathrm{GeV}$ is
\begin{eqnarray}
u_{v}(x),d_{v}(x),g(x) &=&A_{0}\,x^{A_{1}-1}\,(1-x)^{A_{2}}\,
\mathrm{e}^{-A_{3}(1-x)^2\,+\,A_{4}x^{2}} \\
\bar{u}(x)+\bar{d}(x) &=&\textstyle{\frac{1}{2}} \,
A_{0}\,x^{A_{1}-1}\,(1-x)^{A_{2}}\,
\mathrm{e}^{A_{3}x}\,(1+x\mathrm{e}^{A_{4}})^{A_{5}} \\
\bar{d}(x)/\bar{u}(x)
&=&\mathrm{e}^{A_{1}}\,x^{A_{2}-1}\,(1-x)^{A_{3}}\;+\;(1+A_{4}x)\,(1-x)^{A_{5}} \\
s(x)=\bar{s}(x) &=&\textstyle{\frac{1}{2}} \,\kappa \, (\bar{u}(x)+\bar{d}(x)) \\
c(x)=\bar{c}(x)=b(x)=\bar{b}(x) &=& 0
\end{eqnarray}
(Notes regarding these formulas: the parameter $A_1$ is shifted from
the $a_1$ above to make it correspond in definition to the standard
Regge intercept; the parameters $a_3$ and $a_4$ are replaced by
linear combinations $A_3$ and $A_4$ to reduce their correlation in
the fitting by making them control behavior at large and small $x$;
the parameter $A_4$ in $\bar{u}+\bar{d}$ is defined using an
exponential form to ensure positivity of the PDFs.)

For concreteness, we give in Table II the coefficients that
correspond to the central fit CTEQ6.5M. The value of $\kappa $ is
$0.5$; and the strong coupling constant $\alpha _{s}(M_{Z})$ is
$0.118$. We use $M_c = 1.3 \, \mathrm{GeV}$,  $M_b = 4.5 \,
\mathrm{GeV}$.

\begin{tabular}{crrrrrr}
& $A_{0}$ & $A_{1}$ & $A_{2}$ & $A_{3}$ & $A_{4}$ & $A_{5}$  \\
  \hline
          $u_{v}$ & $5.76388$ & $ 0.64416$ & $ 2.31531$ & $ 0.74528$ & $-2.14868$ & $       $ \\
          $d_{v}$ & $3.65786$ & $ 0.62677$ & $ 3.31531$ & $ 0.87977$ & $-2.37338$ & $       $ \\
              $g$ & $0.09974$ & $ 0.22853$ & $ 4.00000$ & $-4.23974$ & $ 8.64169$ & $       $ \\
$\bar{d}+\bar{u}$ & $0.29563$ & $-0.22165$ & $12.09400$ & $ 6.46763$ & $ 4.51075$ & $0.26278$ \\
$\bar{d}/\bar{u}$ & $       $ & $11.49257$ & $ 5.64186$ & $17.00000$ & $19.41872$ & $9.45863$ \\
  \hline
\end{tabular}